\RequirePackage{mathtools}
\RequirePackage[undo-recent-deprecations]{expl3} 
\RequirePackage[2020-02-02]{latexrelease}

\documentclass{jfp1}

\usepackage{amssymb}
\usepackage{array}
\usepackage{ebproof}
\usepackage{multirow}

\title[A Theory of RPC Calculi for Client-Server Model]
      {A Theory of RPC Calculi for Client-Server Model\thanks{  This work was supported by the National Research Foundation of Korea (NRF) grant 
  funded by the Korea government(MSIP) (No. 2017R1A2B4005138).}}

 \author[K. Choi and B-M. Chang]
        {Kwanghoon Choi\\
         Chonnam National University, Gwangju, Republic of Korea\\
         Byeong-Mo Chang\\
         Sookmyung Women's University, Seoul, Republic of Korea\\
         \email{kwanghoon.choi@jnu.ac.kr,chang@sookmyung.ac.kr}}

\jdate{September 2001, update April 2007}
\pubyear{2001}
\pagerange{\pageref{firstpage}--\pageref{lastpage}}
\doi{S0956796801004857}

\newcommand{\rpc}{$\lambda_{rpc}$}
\newcommand{\stateencrpc}{$\lambda_{rpc}^{enc}$}
\newcommand{\statefulrpc}{$\lambda_{rpc}^{state}$}

\newcommand{\stateenccs}{$\lambda_{cs}^{enc}$}
\newcommand{\statefulcs}{$\lambda_{cs}^{state}$}

\newcommand{\client}{\textbf{c}}
\newcommand{\server}{\textbf{s}}

\newcommand{\evalRPC}[3]{#1\Downarrow_{#2}#3}
\newcommand{\evalRPCC}[2]{#1\Downarrow_{\client}#2}
\newcommand{\evalRPCS}[2]{#1\Downarrow_{\server}#2}
\newcommand{\lamL}[3]{\lambda^{#1}#2.#3}
\newcommand{\appL}[3]{#1{\ }^{#2}#3} 
\newcommand{\subst}[2]{\{#1/#2\}}
\newcommand{\llet}[3]{\textsf{let} \ #1 = #2 \ \textsf{in} \ #3}

\newcommand{\textsfReq}{\textsf{req}}
\newcommand{\req}[2]{\textsfReq(#1,#2)}

\newcommand{\textsfCall}{\textsf{call}}
\newcommand{\call}[2]{\textsfCall(#1,#2)}

\newcommand{\textsfRet}{\textsf{ret}}
\newcommand{\ret}[1]{\textsfRet(#1)}

\newcommand{\funL}[1]{\xrightarrow{#1}}    
  
\newcommand{\tyenv}{\Gamma}     
\newcommand{\tyenvExt}[2]{\Gamma\{#1:#2\}}
\newcommand{\typing}[4]{#1\rhd_{#2} #3 : #4}       
\newcommand{\typingBlack}[4]{#1\blacktriangleright_{#2} #3 : #4}  

\newcommand{\loceta}[2]{{#1}\rightsquigarrow{#2}}

\newcommand{\RightarrowEnc}{\Rightarrow^{enc}}
\newcommand{\RightarrowEncStar}{\Rightarrow^{enc*}}
\newcommand{\RightarrowEncPlus}{\Rightarrow^{enc+}}

\newcommand{\runStateEncRPCStar}[2]{$#1 \RightarrowEncStar #2$}

\newcommand{\runStateEncCS}[2]{$#1 \Rightarrow^{enc} #2$}
\newcommand{\runStateEncCSStar}[2]{$#1 \Rightarrow^{enc*} #2$}

\newcommand{\runStatefulCS}[2]{$#1 \Rightarrow^{state} #2$}
\newcommand{\runStatefulCSStar}[2]{$#1 \Rightarrow^{state*} #2$}

\newcommand{\emp}{\epsilon}
\newcommand{\substzsxs}{\{\bar{v}/\bar{z},\overline{w}/\bar{x} \}}

\newcommand{\substXs}{\{\overline{W}/\bar{x} \}}

\newcommand{\LetK}[2]{\textsf{ctx}\ #1 \ #2}
\newcommand{\ccomp}[1]{C[\![#1]\!]}
\newcommand{\scomp}[1]{S[\![#1]\!]}
\newcommand{\vcomp}[1]{V[\![#1]\!]}
\newcommand{\cconv}[1]{CC[\![#1]\!]}

\newcommand{\clo}[2]{clo({#1},{#2})}

\newcommand{\sessionNothing}{\makebox[0.3cm][c]{\scriptsize $nothing$}}
\newcommand{\sessionSomething}{\makebox[0.3cm][c]{\scriptsize $session$}}
\newcommand{\sessionOption}{\makebox[0.3cm][c]{\scriptsize $optSession$}}

\newtheorem{lemma}{Lemma}[section]
\newtheorem{theorem}{Theorem}[section]
\newtheorem{fact}{Fact}[section]
\newtheorem{definition}{Definition}[section]

\begin{document}

\label{firstpage}

\maketitle

\begin{abstract}
	With multi-tier programming languages, programmers can specify the locations of code to run in order to reduce development efforts for the web-based client-server model where programmers write client and server programs separately and test the multiple programs together. 
	The RPC calculus, one of the foundations of those languages by Cooper and Wadler, has the feature of symmetric communication in programmer's writing arbitrarily deep nested client-server interactions. The feature of the calculus is fully implemented by asymmetric communication in trampolined style suitable for the client-server model. 
	However, the existing research only considers a stateless server strategy in which all server states are encoded for transmission to the client so that server states do not need to be stored in the server. It cannot always correctly handle all stateful operations involving disks or databases. 
	To resolve this problem, we first propose new stateful calculi that fully support both symmetric communication from the programmer's viewpoint and asymmetric communication in its implementation using trampolined style. 
	All the existing calculi either provide only the feature of asymmetric communication or propose only symmetric implementation suitable for the peer-to-peer model, rather than the client-server model.
	Second, the method used to design our stateful server strategy is based on a new locative type system which paves the way for a theory of RPC calculi for the client-server model. 
	Besides proposing the new stateful calculi, this theory can improve the existing stateless server strategy to construct new state-encoding calculi that eliminate runtime checks on remote procedure calls present in the existing strategy, and it enables us to design a new mixed strategy that combines the benefits of both kinds of strategies. 
	As far as we know, there are no typed multi-tier calculi that offer programmers the feature of symmetric communication with the implementation of asymmetric communication under the three strategies together. 
\end{abstract}

\tableofcontents

\section{Introduction}
\label{sec:intro}

	Modern computing environments such as web systems involve programming not a single machine but several distributed machines together. For example, a web system basically consists of a web server that accesses databases and a web client that provides user interfaces, and they are connected by a network. Programmers have to develop two individual programs separately for the two machines, which increases the programmer's burden. Further, they need to test the two programs together, which is more complex than they do with one program on a single machine.

	The programmer's task can be alleviated by the use of {\it multi-tier programming languages} \cite{Cooper:2006:LWP:1777707.1777724,Cooper:2009:RC:1599410.1599439,MurphyVII:2004:SML:1018438.1021865,Murphy:2008:MTM:1467784,Neubauer:2005:SPM:1040305.1040324,Rastogi:2014:WPL:2650286.2650775, Serrano2006, Serrano:2016:GH:3022670.2951916, Balat2006, Chlipala2015}. Such languages are equipped with the locations feature, which enables description of the location to describe where a specified part of the code should execute. Using this feature, programmers develop a single program that can be used freely in different locations such as the server and the client in one programming language. Then compilers will automatically generate separate programs for each location, guaranteeing communication integrity among the differently located programs. 

	Notably, RPC calculus \cite{Cooper:2009:RC:1599410.1599439} offers the feature of  symmetric communication between client and server.
After adding to each function location annotations specifying where it must run, programmers can write arbitrarily deep nested client-server interactions by applying only the standard functions. Their compilation method will  automatically map the deep interactions onto flat request-response interactions on the web system. 
	The RPC calculus is the foundation of a practical multi-tier web programming language called Links \cite{Cooper:2006:LWP:1777707.1777724}. 
	
	The existing target language of the RPC calculus, called the CS calculus, where a client and a server run separate programs, was designed for the server to maintain no session with individual clients. All server states during client-server interactions are appropriately encoded in the server for transmission to the client so that server states do not need to be stored in the server.
	The CS calculus uses asymmetric communication, which is almost free in the client-server model, and supports symmetric communication from the programmer's viewpoint using {\it trampolined style} \cite{Ganz:1999:TS:317636.317779}. 
	Thus, the implementation of the RPC calculus was aligned  with the well-known {\it RESTful} architecture of web systems, where web services consist of stateless operations. 
	
	However, the existing RPC and CS calculi cannot always correctly handle all stateful operations involving disks or databases. Some server states with disks or databases are not easily serialised, and therefore, when the server calls a client function, for example, between two subsequent database operations, it is not easy to encode all server states left after the client function call. Even when server states are serialisable, passing serialised server states between client and server repeatedly would increase communication overheads, giving rise to efficiency concerns. To address this problem, a stateful server strategy for the RPC calculus is necessary. This looks natural because supporting stateful interactions is common in web systems. For example. Java HttpSession provides a way to identify a user across more than one page request or visit to a web site and to store information about that user. 
	
\begin{figure}[t]
\centering
\includegraphics[width=0.9\textwidth]{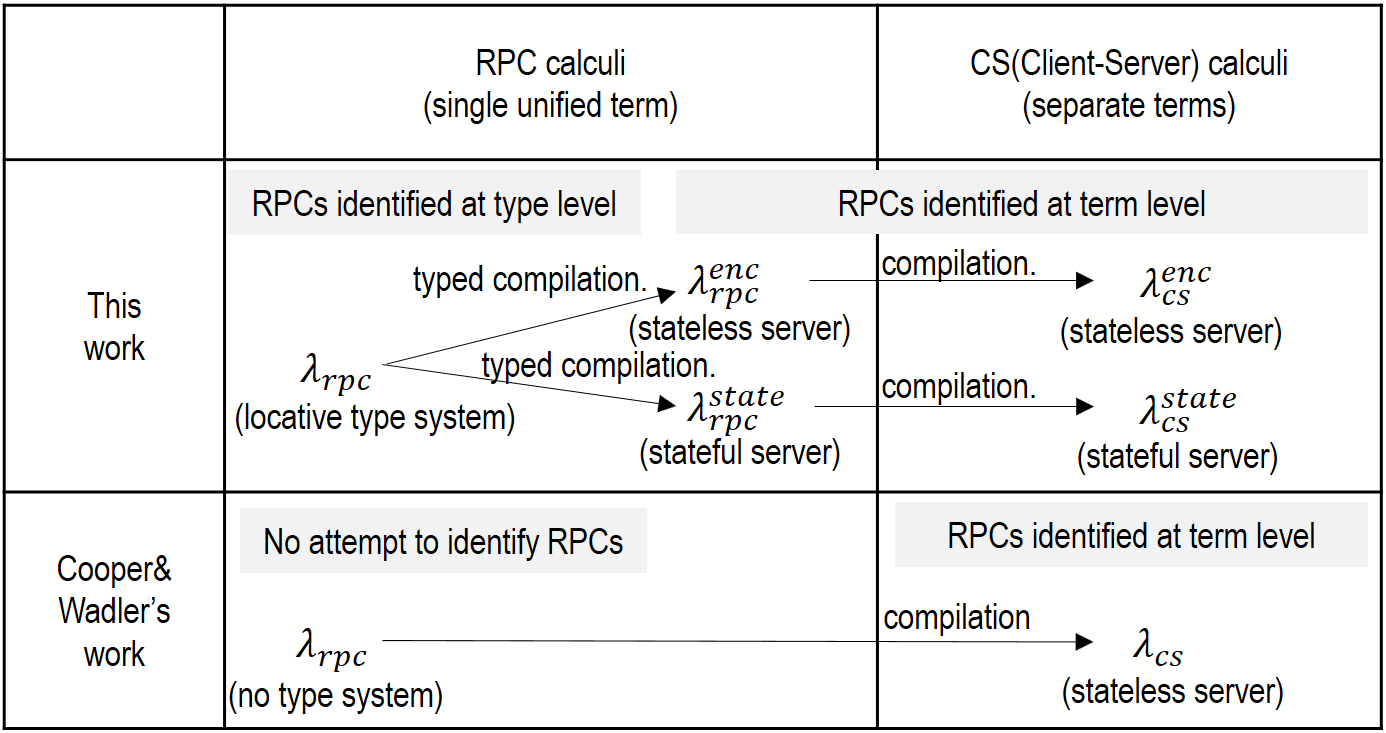}
\caption{Overview of a theory of RPC calculi}
\label{fig:overview}
\end{figure}		
	
	First, we propose new calculi, {\statefulrpc} and {\statefulcs} as shown in Figure \ref{fig:overview}, based on a stateful server strategy to resolve the problems of the existing calculi. The new calculi explicitly represent server states using the runtime stack, as in the conventional programming languages. Most importantly, the new calculi fully support the feature of symmetric communication in the RPC calculus by using trampolined style that is easy to implement in the client-server model. There have been several multi-tier web programming languages adopting stateful strategies, but none of them support both symmetric communication from the programmer's viewpoint and asymmetric communication from the implementor's viewpoint. 
	Hop \cite{Serrano2006, Serrano:2016:GH:3022670.2951916}, Ur/Web \cite{Chlipala2015}, and Eliom \cite{Balat2006,Radanne2017} only provide programmers asymmetric communication where only the client invokes server functions freely but, for the other way around, they must use a special network library, rather than a language construct. For example, Hop offers a web reactive programming library for server-to-client communication. The stateful strategies employed by the three works are different from our strategy fully supporting symmetric communication.
	Lambda5 \cite{MurphyVII:2004:SML:1018438.1021865,Murphy:2008:MTM:1467784} does provide programmers symmetric communication but with distinct syntactic constructs for local and remote functions. The idea of its implementation is similar in that {\it continuations} span multiple worlds such as client and server, but it is based on stateful peer strategies. The semantic rules for all machines in a distributed system are the same, and so there is a gap between the semantic description for the peer-to-peer model and its implementation for the client-server model; to make it fit for the client-server model, they must employ some ad hoc method. Also note that, to make it adopt a stateless server strategy, the semantic rules for sever must be made different from those for client.
	A multi-tier calculus \cite{Neubauer:2005:SPM:1040305.1040324} is also for concurrently running processes, and so it suffers from the same limitation that Lambda5 has. 

	Second, the method used to design our stateful server strategy is based on a new locative type system, and it paves the way for a theory of RPC calculi for the client-server model. Under this theory, besides proposing the new stateful calculi, the theory can improve the existing stateless server strategy to construct new state-encoding calculi, {\stateencrpc} and {\stateenccs} as shown in Figure \ref{fig:overview}, that eliminate runtime checks on remote procedure calls in the existing strategy. The theory also enables us to design a new mixed strategy that combines the benefits of the state-encoding and stateful server strategy. We will elaborate more about our theory as follows.

	A new RPC calculus {\rpc} in the theory is a typed version of the original RPC calculus with locative function types $\tau \funL{a}\tau'$ of functions that should run at location $a$. Typing with locative types can distinguish remote function applications from local ones at the type level, rather than at the term level by distinct constructs as in all the existing calculi except the RPC calculus. 
	Using this type information, we are able to design a new type-directed compilation method. The new method is simpler than the existing untyped method because it has two kinds of compilation rules, one for client and the other for server, rather than eight kinds of rules, three for client and five for server, in the existing method. Also, the use of type information eliminates runtime checks on locations, which is present in the existing untyped compilation method. For example, due to the absence of location information in function application terms, both a local function and a remote one can be invoked through the same function application term, e.g. $f \ M$. So, $f$ is needed to check in runtime if it is a remote function because an invocation of remote functions is implemented differently from an invocation of local functions. With the locative types, either $f$ is always a local function or it is always a remote function.
	
	Our theory not only allows us to enjoy the benefits of the simple type-directed compilation method and the elimination of the runtime checks on locations, but it also provides a framework to compare and to combine the two kinds of strategies. 
	Under the same theory, it is easy to compare the two server strategies in terms of session management. In the state-encoding server strategy, one session corresponds to a single request-response interaction, whereas one session in the stateful server strategy can span multiple request-response interactions. A formal description of the comparison will be presented. 
	Because our theory employs a typed approach, it is also possible to design a mixed strategy where the state-encoding server strategy is basically used to reduce the server resource consumption but we switch to the stateful server strategy when necessary. 
	This idea of a mixed strategy can be realized by adopting the monadic encapsulation of state \cite{Launchbury1994,Timany:2017}. Using the notion, stateful computations are encapsulated using monads, and they can be separated from the purity of functional language. In the design, we use the stateful calculi for the phase that uses stateful operations separated by the monadic type system, and we use the state-encoding calculi for the other phase that uses purely functional operations.  
	
		For the evaluation, we have implemented a prototype compiler of {\rpc} into two kinds of calculi with a locative type inference algorithm, and have implemented two evaluators running a client and a server communicating with the HTTP protocol\footnote{https://github.com/kwanghoon/rpccalculi}. 	

	Figure \ref{fig:overview} shows an overview of our theory. As far as we know, there are no typed locative calculi with the feature of symmetric communication that are implemented with asymmetric communication using a stateless server strategy, a stateful server strategy, and a mixed strategy together. 
		
The contributions of this paper are as follows: 
\begin{itemize}
\item We present a locative type system for the RPC calculus {\rpc} and prove its type soundness, guaranteeing that the locative information is preserved under the evaluation. 
\item We design the two RPC calculi {\stateencrpc} and {\statefulrpc} with two semantically correct compilation methods of {\rpc} into {\stateencrpc} and {\statefulrpc}, respectively. 
\item We also design two CS calculi {\stateenccs} and {\statefulcs} with two semantically correct compilation methods of {\stateencrpc} into {\stateenccs} and of {\statefulrpc} and {\statefulcs}. 
\item We implement our stateless and stateful calculi to show their effectiveness by  a location-type inference algorithm, two compilers, and two HTTP-based evaluators. 
\item We formally compare the characteristics of session management of {\stateenccs} with those of {\statefulcs} by extending the semantic rules with session annotations. 
\item We propose a method to design a mixed strategy of the two strategies by extending the RPC calculus with monadic encapsulation of state. 
\end{itemize}

	Section \ref{sec:rpc} introduces the RPC calculus {\rpc} with a locative type system and proves the type soundness of the calculus. 
	Section \ref{sec:stateenccalculi} proposes two calculi {\stateencrpc} and {\stateenccs} as a new formulation of the server state-encoding strategy to implement the {\rpc} calculus. 
	Section \ref{sec:statefulcalculi} extends the new formulation to two stateful calculi {\statefulrpc} and {\statefulcs} to implement the {\rpc} calculus with the stateful server strategy. 
	Section \ref{sec:relatedwork} discusses related work.
	Section \ref{sec:conclusion} concludes the paper. 
	In the appendix, all proofs of theorems introduced in the sections are available. 

\section{The RPC Calculus and Its Locative Type System} 
\label{sec:rpc}

Let us review the RPC calculus \rpc, proposed in \cite{Cooper:2009:RC:1599410.1599439}. This is an ordinary call-by-value $\lambda$-calculus with location annotations on $\lambda$-abstractions. The annotations tell the locations where the $\lambda$-abstractions execute. This calculus was designed as the foundation of a practical multi-tier web programming language. The client-server model is assumed in the calculus, and so the location annotations are either {\client} denoting client  or {\server} denoting  server. The syntax of {\rpc} is thus defined as in \mbox{Figure \ref{fig:rpc}}.  

\setlength{\tabcolsep}{3pt}

\begin{figure}[t]
\begin{tabular}{l l l c l c l }
\multicolumn{7}{l}{\textbf{Syntax}} \\
  \ \ \ \ \ & Location & $a,b$ 	& $::=$ & $\client$	& $|$		& $ \server$		 \\
            & Term     & $L,M,N$	& $::=$ & $V$		& $ |$	& $L \ M$			 \\
            & Value    & $V,W$	& $::=$ & $x$		& $ |$	& $ \lambda^a x.N$	   
\\[0.1cm]
\multicolumn{7}{l}{\textbf{Semantics}} \\
  & \multicolumn{6}{c}{
  	\mbox{
       \begin{prooftree}
       		\infer0[(Value)]{ \evalRPC{V}{a}{V} }
	\end{prooftree}
	\ \ \ \ \ 
	\begin{prooftree}
  		\hypo{ \evalRPC{L}{a}{\lamL{b}{x}{N}} }
  		\hypo{ \evalRPC{M}{a}{W} }
  		\hypo{ \evalRPC{N\subst{W}{x}}{b}{V}  }
  		\infer3[(Beta)]{ \evalRPC{L \ M}{a}{V}  }
	\end{prooftree} 
  	}
    } 
\end{tabular}
\\[0.1cm]
\caption{The RPC calculus \rpc}
\label{fig:rpc}
\end{figure}

	In Figure \ref{fig:rpc}, {\rpc} has a big-step operational semantics with evaluation judgements, \[
\evalRPC{M}{a}{V}
\]
denoting the evaluation of a term $M$ in the location $a$ resulting in the value $V$.
In the semantics, $\beta$-reduction is performed in the location annotated on the $\lambda$-abstraction. $N\subst{W}{x}$ is an ordinary substitution of $W$ for $x$ in $N$. 

\begin{figure}[h]
\centering
\includegraphics[width=0.45\textwidth]{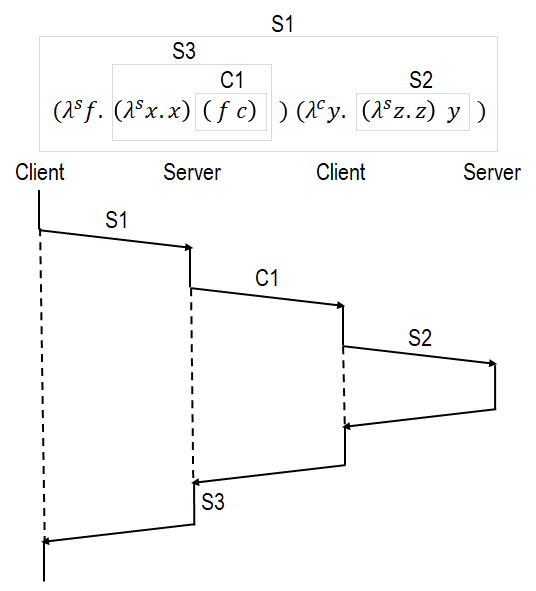}
\caption{Evaluation flow in {\rpc} starting from the client (Labels C1, S1, S2, and S3  indicate the locations of the applications in the labelled boxes)}
\label{fig:rpcexample}
\end{figure}

	In {\rpc}, the evaluation of a term starts from client. An example is:
\[
(\lamL{\server}{f}{ \ (\lamL{\server}{x}{x}) \ (f \ \ c) }) \ \ (\lamL{\client}{y}{  \ (\lamL{\server}{z}{z}) \ \ y   })
\]
Figure \ref{fig:rpcexample} depicts the flow between the client and server for the example. 
The client first requests the server to invoke a function $(\lamL{\server}{f}{\cdots})$ at the application in a box named S1. During the evaluation of the body of the server function, the server calls a client function $f$ in C1, which is $(\lamL{\client}{y}{\cdots})$, with some constant $c$ as an argument. This client function again requests the server to invoke an identity function $(\lamL{\server}{z}{z})$ in S2. Finally, a server function $(\lamL{\server}{x}{x})$ in S3 is locally applied to the result of $(f \ c)$.  

\subsection{A Locative Type System for {\rpc}}

	We propose a new type system for the untyped RPC calculus {\rpc} where function types 
\[ \tau \funL{a}\tau'
\]
carry location annotations such that $\lambda$-abstractions of the function type run in the location $a$ specified in the type annotation. We advocate the use of types because a typed RPC calculus is useful for compilation, which will be shown later. Figure \ref{fig:rpctysystem} shows type terms and typing rules. Here, $base$ denotes fundamental types such as int.

\begin{figure}[h]
\begin{tabular}{l l l l l l l }
\multicolumn{7}{l}{\textbf{Types}} \\
   \ \ \ \ \ & Type & $\tau$ & $::=$ & $base$ & $ | $ & $ \tau\funL{a}\tau$  \\[0.1cm]
\multicolumn{7}{l}{\textbf{Typing Rules}} \\
  & \multicolumn{6}{c}{  
  \mbox{
  	\begin{prooftree}
		\hypo{  \tyenv(x)=\tau }
		\infer[left label=(T-Var)]1{ \typing{\tyenv}{a}{x}{\tau} }
	\end{prooftree}
	\ \ \ \ \ 
	\begin{prooftree}
		\hypo{ \typing{\tyenvExt{x}{\tau}}{b}{M}{\tau'} }
		\infer[left label=(T-Lam)]1{ \typing{\tyenv}{a}{\lamL{b}{x}{M}}{\tau\funL{b}\tau'} } 
	\end{prooftree}
	\ \ \ \ \ 
	\begin{prooftree}
		\hypo{ \typing{\tyenv}{a}{L}{\tau\funL{a}\tau'} }
		\hypo{ \typing{\tyenv}{a}{M}{\tau}  }
		\infer[left label=(T-App)]2{ \typing{\tyenv}{a}{L \ M}{\tau'} }
	\end{prooftree}
    } 
    }
\\[0.2cm]
 & 
 \multicolumn{6}{c}{  
 \mbox{
	\begin{prooftree}
		\hypo{  \typing{\tyenv}{\client}{L}{\tau\funL{\server}\tau' } }
		\hypo{  \typing{\tyenv}{\client}{M}{\tau} }
		\infer[left label=(T-Req)]2{ \typing{\tyenv}{\client}{L \ M}{\tau'}   }
	\end{prooftree}
	 	\ \ \ \ \ 
	\begin{prooftree}
		\hypo{ \typing{\tyenv}{\server}{L}{\tau\funL{\client}\tau'}  }
		\hypo{ \typing{\tyenv}{\server}{M}{\tau}  }
		\infer[left label=(T-Call)]2{  \typing{\tyenv}{\server}{L \ M}{\tau'} }
	\end{prooftree}
	}
    }
\end{tabular}
\\[0.1cm]
\caption{A Locative Type System for \rpc}
\label{fig:rpctysystem}
\end{figure}

	This type system uses typing judgements of the form $\typing{\tyenv}{a}{M}{\tau}$, which denotes that a term $M$ has type $\tau$ under a type environment $\tyenv$ at location  $a$. 
By (T-Lam), a location specified in the $\lambda$-abstraction is defined to be carried by the function type, e.g. as \mbox{$\typing{\tyenv}{\client}{\lamL{\server}{z}{z}}{\tau\funL{\server}\tau}$} with location $\server$. (T-Var) is defined as usual.
	There are three typing rules for applications classified by where the applications execute and where the functions of the applications execute. (T-App) pertains to local applications. The location where the functions of the application execute is the same as the location where the application executes. (T-Req) and (T-Call) are about remote applications. (T-Req) is a typing for applications of a server function requested by the client, whereas (T-Call) is a typing for applications of a client function called by server. Here is a typing derivation example: 
\[
\begin{prooftree}
	\hypo{ \typing{\tyenv}{\client}{\lamL{\server}{z}{z}}{\tau\funL{\server}\tau}  }
	\hypo{ \typing{\tyenv}{\client}{y}{\tau}  }
	\infer[left label=(T-Req)]2{ \typing{\tyenv}{\client}{   (\lamL{\server}{z}{z}) \ y }{\tau}  }
\end{prooftree}
\]

	The example {\rpc} term in Figure \ref{fig:rpcexample} can be typed under our type system using (T-Req) for the two function applications in the boxes S1 and S2, respectively, (T-App) for the function application in S3, and (T-Call) for the function application in C1. 
	
	For convenience, the typing derivation can be represented in an extended syntax as
\[
\appL{(\lamL{\server}{f}{ \ \appL{(\lamL{\server}{x}{x})}{\server}{(\appL{f}{\client}{c})} })}{\server}{(\lamL{\client}{y}{  \ \appL{(\lamL{\server}{z}{z})}{\server}{y}   })}
\]
where $f$ has type $base\funL{\client}base$ and all of $x$, $y$, $z$, and $c$ have type $base$.
	In the extended syntax, each application term gets a location annotation. $\appL{L}{a}{M}$ at location $a$ is for (T-App), $\appL{L}{\server}{M}$ in the client is for (T-Req), and $\appL{L}{\client}{M}$ in the server is for (T-Call). Then the extended syntax of the {\rpc} terms will uniquely determine typing derivations under location contexts. This representation will be used in our compilation later. 

	The proposed locative type system has type soundness in that location information is preserved under the evaluation in {\rpc}, as is proved by the following theorem. 

\begin{theorem}[Type Soundness for \rpc] If \ $\typing{\tyenv}{a}{M}{\tau}$ and $\evalRPC{M}{a}{V}$, then $\typing{\tyenv}{a}{V}{\tau}$.
\end{theorem}

	Note that the typed RPC calculus is a strict subset of the untyped calculus. There are two kinds of untypeable terms. One kind is due to simple type incompatibility in which the base type is not compatible with any function types, which does not interest us much. We could filter out such incompatible terms with the simply typed system. The other kind of untyped terms are due to location incompatibility. Because we refine function types with location information, two lambda terms that would have the same function type under the simply typed system can now have different types. Consider an application term $(\lamL{a}{f}{\cdots}) \ M$ where $f$ has type $base\funL{c}base$ and $M$ has type $base \funL{s} base$. Obviously, this term cannot be typed under the type system for {\rpc}. 
	However, it is not difficult to make it typed without changing the computation by a transformation of $M$ into $\lamL{c}{x}{M \ x}$, which now has the same type as $f$ and the {\it same} computation\footnote{In a calculus with effects, the transformation may delay or eliminate effects because it transforms expressions with effects into values without them. In this situation, the transformation must involve more, for example, $\llet{y}{M}{\lamL{c}{x}{y \ x}}$ to have the same effects.}. This idea can be generalised as a transformation $[\![M]\!]^{\loceta{\tau}{\tau'}}$ where a term $M$ of type $\tau$ is transformed into $\tau'$ and $\tau$ becomes identical to $\tau'$ on the removal of all location annotations in the types.
\[
\mbox{
\begin{tabular}{l c l}
$[\![M]\!]^{\loceta{\tau}{\tau}}$ & $=$ & $M$ \\
$[\![M]\!]^{\loceta{\tau_1\funL{a}\tau_2 \ }{\ \tau_3\funL{b}\tau_4}}$ & $=$ & 
  $\lamL{b}{x}{[\![ M \ \ [\![x]\!]^{\loceta{\tau_3}{\tau_1}} ]\!]^{\loceta{\tau_2}{\tau_4}}}$ \\
\end{tabular}
}
\]

	The above analysis establishes that every typed term in the simply typed system with arbitrary location annotations can be made typed under our locative type system for {\rpc} immediately or by the computation-preserving transformation.
	Our location-typed approach can thus deal with all terms in the existing location-untyped approach.
	
	We have implemented a unification-based type inference algorithm for our locative type system.  Although polymorphic types are beyond the scope of this research, we believe that it is possible to extend our type system with them. 

\section{Locative Calculi Encoding Server States }
\label{sec:stateenccalculi}

	We present state-encoding calculi for the client-server model, following the stateless server strategy in \cite{Cooper:2009:RC:1599410.1599439}. Under this strategy, when the server transfers control to the client, all the server states are encoded into a term so that the client can get the server to begin evaluating the term later. Unlike the original formulation, our calculi are new in that they are based on location information at the type level . This feature improves the original calculus, as we will see. 
	
\begin{figure}[t]
\centering
\includegraphics[width=0.7\textwidth]{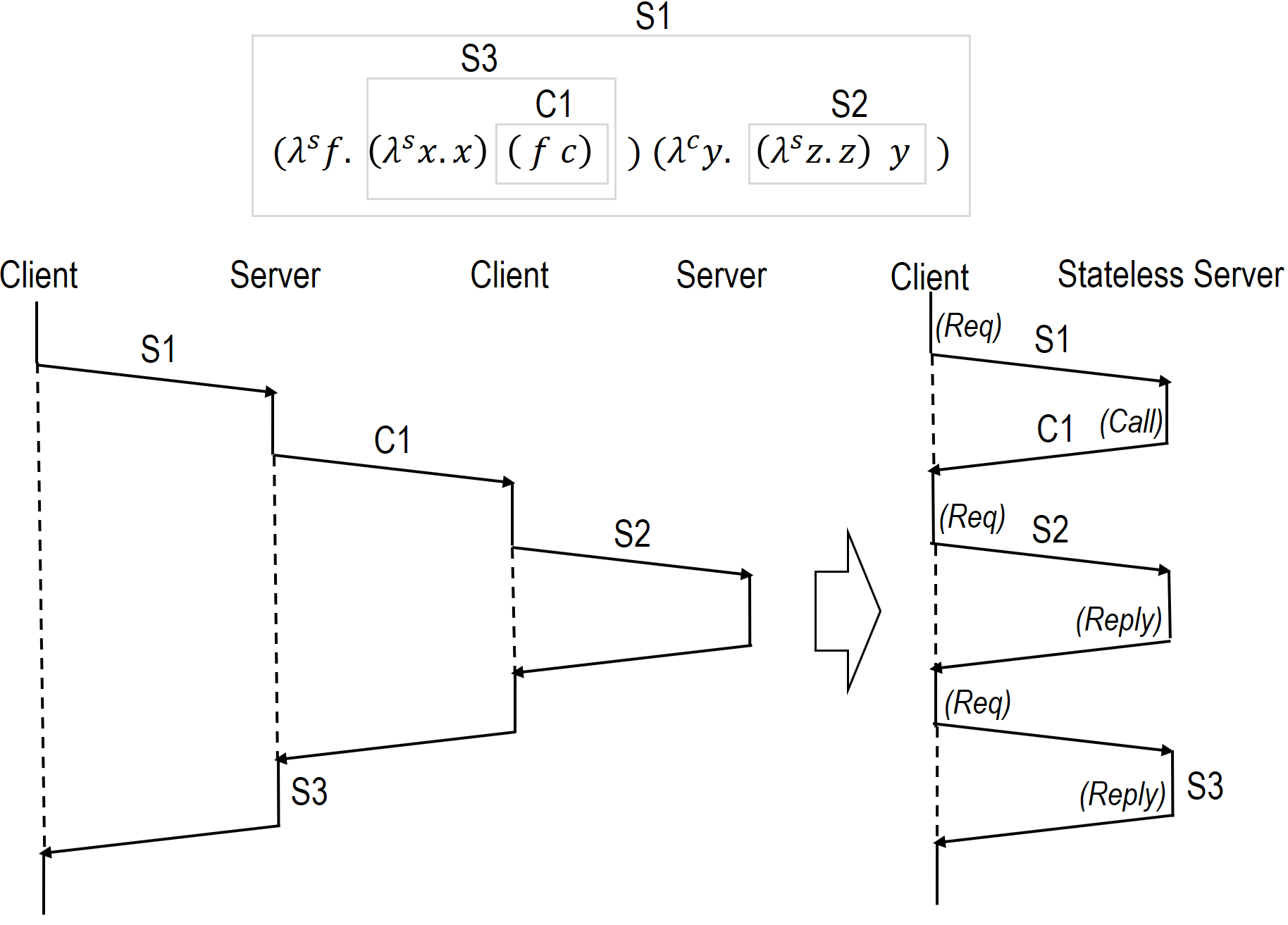}
\caption{Mapping of arbitrarily deep client-server interactions in {\rpc} onto flat request-responses interactions in {\stateencrpc}}
\label{fig:rpcVSstateless}
\end{figure}

	Figure \ref{fig:rpcVSstateless} depicts the basic idea of the stateless server strategy on how to map every deep nesting of client-server interactions (on the left) into a series of request-response interactions between the client and the server (on the right). Each solid line indicates an active thread of control, and each dashed line indicates a stack frame which is waiting for a function to return. 
	To support this mapping on the stateless server, we introduce the notion of {\it continuation} \cite{Flanagan:1993:ECC:155090.155113} to encode the server control to continue. For example, 
	on invoking the client function in (C1), a continuation for (S3) is encoded into a term, and the encoded term is sent to the client. After the evaluation of (C1), the client requests the server to evaluate with the term to continue.
	Accordingly, there are two kinds of responses, as shown in Figure \ref{fig:rpcVSstateless}: one kind of response marked by (Call) involves the return of some encoded continuation, and the other kind marked by (Reply) simply returns values. 

	Under the stateless server strategy, it is assumed that every continuation carried by (Call)-type responses can be serialised so as to return to the client through the network. When this assumption does not hold, the continuation-based translation will not work. This observation leads to an alternative {\it stateful server} strategy in Section \ref{sec:statefulcalculi}, which requires no such assumption. In this section, we develop our theory under the assumption of a serialisable continuation. 

	Now we introduce a new (state-encoding) RPC calculus {\stateencrpc} where local function applications are distinguished from remote function applications explicitly by the new term constructs {\textsfReq} and {\textsfCall}. 

\subsection{A State-Encoding RPC Calculus \stateencrpc}
\label{sec:stateencrpc}
	We first explore the state-encoding RPC calculus informally with an example, and then we will present its formal semantics in the next section.
	In {\stateencrpc}, there are three kinds of function applications: a local one $V_f(\overline{W})$ and two remote ones, $\req{V_f}{\overline{W}}$ and $\call{V_f}{\overline{W}}$. In the local function application, $V_f$ is a local function. $\overline{W}$ denotes a sequence of values that are its arguments. The terms $\req{V_f}{\overline{W}}$ are used only in the client to transfer control to the server and then to invoke a server function $V_f$. The terms $\call{V_f}{\overline{W}}$ are used only in the server to do the converse with a client function $V_f$. 
	
	Consider again the {\rpc} example term:
\[
(\lamL{\server}{f}{ \ (\lamL{\server}{x}{x}) \ (f \ c) }) \ \ \ \ \ (\lamL{\client}{y}{  \ (\lamL{\server}{z}{z}) \ y   })
\]
whose evaluation begins in the client. This {\rpc} term can be implemented under the stateless server strategy by a {\stateencrpc} term as
\[
\llet{r}{\req{M_1}{(M_2, ID)}}{r}
\]
where $M_1$ and $M_2$ implement the outermost function $(\lamL{\server}{f}{(\lamL{\server}{x}{x}) \ (f \ c) })$ and its argument $(\lamL{\client}{y}{  \ (\lamL{\server}{z}{z}) \ y   })$, respectively. We used {\textsfReq} to invoke the server function in the client. As previously explained, the notion of continuation was used to encode server states. In the stateless server strategy, server functions are implemented in continuation-passing style (CPS), whereas client functions are implemented in direct style \cite{Flanagan:1993:ECC:155090.155113}. $M_1$ is a CPS function converted from the server function. $M_2$ as a normal argument is converted from the original argument, and  $ID$ as a continuation argument denotes the identity continuation $\lamL{\server}{w}{w}$. 

	Now let us examine $M_1$: 
\[
\mbox{
\begin{tabular}{l}
$M_1 \triangleq \lamL{\server}{(f,k)}{ 
	\call{\ \ \ \lamL{\client}{x}{\llet{y}{f(x)}{\req{ cont }{y}}}}{\ \ \ [\![ c ]\!] \ \ \ }
	 }$ 
\end{tabular}
}
\]
where $cont$ is $\lamL{\server}{w}{[\![ \lamL{\server}{x}{x}  ]\!] (w,k)}$, 
and $[\![ M  ]\!]$ represents an implementation of the {\rpc} term $M$ for the server. 
	Here, $f$ is the client function to be invoked, and $k$ is a continuation, which is $ID$ in the example. The body of $M_1$ is in the context of the server. 
	We use {\textsfCall} to invoke the client function `$\lamL{\client}{x}{\cdots}$' in $M_1$: it transfers control to the client to evaluate $ f \ [\![ c ]\!]$ by `$\llet{y}{f(x)}{\cdots}$' above, and then it returns control to the server by {\textsfReq} with the continuation $cont$ representing $(\lamL{\server}{x}{x}) \ [-]$ by `$\cdots\ \textsf{in} \ \req{cont}{y}$' above. 
	The client function in {\textsfCall} thus supports the behaviour of {\it commuting from the server to the client}, where $cont$ is regarded as an encoding of the server state. We call such a  client function a {\it commuting function}, and our compiler will thus compose commuting functions. 

	$M_2$ has the following simple structure:
\[
\mbox{
\begin{tabular}{l l}
$M_2 \triangleq \lamL{\client}{y}{}$ &
 $\llet{ \ \ d}{ \lamL{\server}{(z,k)}{k(z)} \ \ }{}$ \\
                                     &
 $\llet{ \ \ r}{\req{d}{(y,ID)} \ \ }{r}$
\end{tabular}
}
\]
where $d$ is a CPS function converted from $\lamL{\server}{z}{z}$, and the client invokes this (server) function with its arguments $y$ and $ID$.

\begin{figure*}[t]
\begin{tabular}{l}
\begin{tabular}{l l}
\multicolumn{2}{l}{\textbf{Syntax}} \\
&
\begin{tabular}{l l l c l }
  & Term & $M$	    & $::=$  & 
  $V \ | \ V_f(\overline{W}) \ | \ \req{V_f}{\overline{W}}
  		\ | \ \call{V_f}{\overline{W}} \ | \ \llet{x}{M}{M}$ \\ 
  & Value & $V,W$  & $::=$  & $x \ \ | \ \ \lamL{a}{\bar{x}}{M}$ \\ 
  & Client context  & $\Pi$  & $::=$  & $\LetK{x}{M}$    \\
  & Server context & $\Delta$  & $::=$  & $\emp$    
\end{tabular}
\end{tabular}
\\
\begin{tabular}{l l}
\multicolumn{2}{l}{\textbf{Semantics}} 
\\
&
\begin{tabular}{l l l l l l } 
 \multicolumn{6}{l }{ 
     \framebox{
     	\begin{minipage}{0.94\textwidth}
		Client: \ \ \ \runStateEncCS{M \ | \ \emp}{M' \ | \ \emp} \ \ \ or \ \  \ 
     			\runStateEncCS{M \ | \ \emp}{\Pi \ | \ M'}  
	\end{minipage} }
} 
\\[0.2cm]
  \ \ \ (AppC)
 &	
 \multicolumn{5}{l}{
 	$\llet{y}{(\lamL{\client}{\bar{x}}{M_0})(\overline{W})}{M} \ | \ \emp
 	\ \ \Rightarrow^{enc} \ \ 
 	\llet{y}{M_0\substXs}{M} \ | \ \emp$}
 \\[0.1cm] 
  \ \ \ (Req)
 &
 \multicolumn{5}{l}{
 $\llet{x}{\req{\lamL{\server}{\bar{x}}{M_0}}{\overline{W}}}{M} \ | \ \emp
 \ \ \Rightarrow^{enc} \ \ 
 \LetK{x}{M} \  | \ (\lamL{\server}{\bar{x}}{M_0})(\overline{W})$}
 \\[0.1cm] 
 \ \ \ (ValC)
 &
 \multicolumn{5}{l}{
 $\llet{x}{V}{M} \ | \ \emp
 \ \ \Rightarrow^{enc} \ \ 
 M\subst{V}{x} \ | \ \emp$}
 \\[0.1cm]
 \ \ \ (LetC)
 &
 \multicolumn{5}{l}{
 $\llet{x}{  (\llet{y}{M_1}{M_2})  }{M} \ | \ \emp
 \ \ \Rightarrow^{enc} \ \ 
 \llet{y}{M_1}{(\llet{x}{M_2}{M})} \ | \ \emp$}
 \\[0.1cm] 
\end{tabular}
\\
&
\begin{tabular}{l l l l l l }
 \multicolumn{6}{l}{ 
     \framebox{
     	     	\begin{minipage}{0.94\textwidth}
			Server: \ \ \ \runStateEncCS{\Pi \ | \ M}{\Pi \ | \ M'} \ \ \ or \ \  \ 
     					\runStateEncCS{\Pi \ | \ M}{M' \ | \ \emp}
		\end{minipage} }
 } 
 \\[0.2cm]
 \ \ \ (AppS) 
 &
 \multicolumn{5}{l}{
 $\Pi \ | \ (\lamL{\server}{\bar{x}}{M_0})(\overline{W})
 \ \ \Rightarrow^{enc} \ \ 
 \Pi \ | \  M_0\substXs$}
 \\[0.1cm]
 \ \ \ (Call)
 &
 \multicolumn{5}{l}{
 $\LetK{x}{M} \ | \ \call{\lamL{\client}{\bar{x}}{M_0}}{\overline{W}}
 \ \ \Rightarrow^{enc} \ \ 
 \llet{x}{(\lamL{\client}{\bar{x}}{M_0})(\overline{W})}{M} \ | \ \emp$}
 \\[0.1cm]
 \ \ \ (Reply)
 &
 \multicolumn{5}{l}{
 $\LetK{x}{M} \ | \ V
 \ \	\Rightarrow^{enc} \ \ 
 \llet{x}{V}{M} \ | \ \emp$}
\end{tabular}
\end{tabular}
\end{tabular}
\\[0.2cm]
\caption{A state-encoding client-server calculus \stateencrpc}
\label{fig:stateencrpc}
\end{figure*}

\subsection{The Formal Semantics of \stateencrpc}
\label{sec:semanticsencrpc}

	This section first introduces the syntax and semantics for the state-encoding RPC calculus, as shown in Figure \ref{fig:stateencrpc}. 
	The syntax of $M$ terms resembles A-normal form \cite{Flanagan:1993:ECC:155090.155113}. 
	Values are a variable or a location-annotated function and are denoted by $V$ or $W$. Note that $\bar{x}$ denotes a sequence of variables, and $\overline{V}$ denotes a sequence of values.
	There are two kinds of remote function calls: $\call{V_f}{\overline{W}}$ for invocation of a client function $V_f$ in the server with arguments $\overline{W}$, and $\req{V_f}{\overline{W}}$ for invocation of a server function in the client.
	The let constructs $\llet{x}{M_1}{M_2}$ bind $x$ to some intermediate value from the evaluation of $M_1$, and continue to evaluate $M_2$.
	
	The client context $\Pi$, which is $\LetK{x}{M}$, waits for a value to bind for $x$ and to evaluate $M$. The server context $\Delta$ for the state-encoding RPC calculus is always empty. 

	In the semantics, $client \ | \ server$ is the notation for snapshots in a client-server based distributed system. We call such a snapshot a {\it configuration}. There are two kinds of configurations in this calculus:
\begin{itemize}
\item $M \ | \ \emp$ for the client to evaluate a term $M$ with the empty server context ($\emp$)
\item $\Pi \ | \ M$ for the server to evaluate a term $M$ with a pending client context $\Pi$ 
\end{itemize}

	The semantics is described as a relation on configurations as 
	\[ Conf \ \ \Rightarrow^{enc} \ \ Conf'
	\]
	Local evaluation is in the form of \runStateEncCS{M \ | \ \emp}{M' \ | \ \emp} for the client and \runStateEncCS{\Pi \ | \ M}{\Pi \ | \ M'} for the server. When control moves from the client to the server, the relevant semantic rule will have the form \runStateEncCS{M \ | \ \emp}{\Pi \ | \ M'}. For the opposite direction, it is \runStateEncCS{\Pi \ | \ M}{M' \ | \ \emp}. The notation $R ^{+}$ and $R ^{*}$ for a relation $R$ are defined as usual.

	In the semantics, the evaluation of a term $M$ begins in the client with the empty server state and normally finishes  when it reaches the value $V$, as
	$M \ | \ \emp \ \Rightarrow^{enc *} \ V \ | \ \emp 
	$, meaning that $M$ evaluates to $V$.
	 
	For evaluation in the client, (AppC) performs a client function application. $M\substXs$ denotes an ordinary parallel substitution of $\overline{W}$ for $\bar{x}$ in $M$. (ValC) binds an intermediate value with a let-variable and continues to evaluate a term in the let body. (LetC) takes a nested let binding out for evaluation. With (Req), the client can request the server to execute a server function application, leaving the client context. 

	The server-side evaluation, which is initiated by (Req) in the client, continues with  (AppS). It is reminiscent of (AppC) for the client. The evaluation in the server will reach either a value $V$ or a call $\call{f'}{\overline{W'}}$. For the case of a value, (Reply) returns the value to the client context $\LetK{x}{M}$ as  $\llet{x}{V}{M}$. For the case of a call, (Call) moves the client function call into the client context as $\llet{x}{f'(\overline{W'})}{M}$. In any case, the server state will be empty. This is the stateless server strategy.  The typical steps in the server are as follows: 
	\begin{center}
	\begin{tabular}{l l l l l l}
                      & & [Client]                   &                  &[Server] \\                    
         Initiation & & $ \llet{x}{\req{f}{\overline{W}}}{M}$ & $ \  |  \ $ & $ \emp$ & by (Req) \\
  & $\Rightarrow^{enc}$  & $\LetK{x}{M}$ & $  \ | \ $ & $ f(\overline{W})$ &  \\[0.1cm]
  either & $\Rightarrow^{enc \ +}$  & $\LetK{x}{M}$    & $  \ | \ $ & $V$ & by (Reply) \\
  & $\Rightarrow^{enc}$      & $\llet{x}{V}{M}$ & $  \ | \ $ & $\emp$  &   \\[0.1cm]
  or & $\Rightarrow^{enc \ +}$  & $\LetK{x}{M}$    & $  \ | \ $ & $\call{f'}{\overline{W'}}$ & by (Call) \\
  & $\Rightarrow^{enc}$      & $\llet{x}{f'(\overline{W'})}{M}$ & $  \ | \ $ & $\emp$  &   \\
	\end{tabular} 
	\end{center}

	The proposed semantic rules are defined in a minimal way enough for {\stateencrpc} to be an intermediate calculus for compilation of {\rpc}, as will be seen. The semantic rules could be easily extended to cover all syntactic terms including, for example, local applications without let in (AppC) or remote applications without let in (Req), but this extension is not essential for our purpose. In the server side, the intermediate calculus for CPS conversion has only to support local applications, $\textsfCall$, and values, all without let. 

	Note that the evaluation of $\textsfCall$ in the semantics is allowed only in tail positions because the server state right after the call must be empty. To map {\rpc} terms without such a restriction into {\stateencrpc} terms with it, compiler support is required as will be explained soon. 

\subsection{A Typed Compilation of {\rpc} into {\stateencrpc}}
\label{sec:compforstateenccs}

	Now we are ready to present a typed compilation of the RPC calculus into the state-encoding RPC calculus. Figure \ref{fig:compforstateencrpc} shows our compilation rules. 

	The compilation rules for client terms are described as $\ccomp{M_{rpc}}=M_{rpc}^{enc}$.	Here, $M_{rpc}$ actually denotes a typing derivation for a source term in the RPC calculus, rather than the source term itself. Such a typing derivation provides each application subterm location information where the subterm should be evaluated, as explained in Section \ref{sec:rpc}.
	$M_{rpc}^{enc}$ denotes a compiled target term. 		
	
	(VarC) compiles a variable term simply into a variable. (LamCC) compiles each client function using direct style. (LamCS) compiles each server function using CPS conversion.  The reason for using the conversion will be explained later.
	(AppCC) is a compilation rule for a local client function application typed under the typing rule (T-App), denoted by $\appL{L}{\client}{M}$. It is a conventional compilation rule that compiles $L$ into a term generating a client function, compiles $M$ into another term generating an argument, and ends with a local application term. 
	(AppCS) compiles a remote server function application typed under the typing rule (T-Req), denoted by $\appL{L}{\server}{M}$. It is almost the same as (AppCC). The only difference is the replacement of a local application term $(f \ x)$ with a remote application term  
$
	\req{f}{\ (x,\lamL{\server}{y}{y})}
$.
Our compilation rules ensure that $f$ is a CPS-converted function. Therefore, it will take not only a function argument $x$ but also a continuation. The identity continuation ($\lamL{\server}{x}{x}$) is given to the remote application term because it begins the evaluation of a CPS term. 

	(AppCC) and (AppCS) introduce the extra layer of `$\llet{r}{\cdots}{r}$' surrounding local and remote applications to adapt our compilation to the minimal semantic rules (AppC) and (Req), respectively. 

\begin{figure*}
\begin{tabular}{l l l c l }
 \multicolumn{5}{l}{
  \framebox{Client: \ $\ccomp{M_{rpc}} \ = \ M_{rpc}^{enc}$}
} 
\\
 & (VarC)  & $\ccomp{x}$ & $=$ & $x$
\\
 & (LamCC) & $\ccomp{\lamL{\client}{x}{M}}$ & $=$ & $\lamL{\client}{x}{\ccomp{M}}$
  \\
 & (LamCS) & $\ccomp{\lamL{\server}{x}{M}}$ & $=$ & $\lamL{\server}{(x,k)}{\scomp{M}~k}$
  \\
 & (AppCC) & $\ccomp{\appL{L}{\client}{M}}$ & $=$ & $\llet{f}{\ccomp{L}}{}$ \\
  &                &                                                          &         &  $\llet{x}{\ccomp{M}}{}$ \\
  &                &                                                          &         &  $\llet{r}{f(x)}{r}$
 \\
 & (AppCS) & $\ccomp{\appL{L}{\server}{M}}$ & $=$ & $\llet{f}{\ccomp{L}}{}$ \\
  &                &                                                          &         &  $\llet{x}{\ccomp{M}}{}$ \\
  &                &                                                          &         &  $\llet{r}{\req{f}{(x,\lamL{\server}{y}{y})}}{r}$            
 \\[0.1cm]
\multicolumn{5}{l}{
  \framebox{Server: \ $\scomp{M_{rpc}} \ K \ = \ M_{rpc}^{enc}$}
} 
\\
 & (VarS)  & $\scomp{x} \ K$ & $=$ & $K(x)$
\\
 & (LamSC) & $\scomp{\lamL{\client}{x}{M}} \ K$ & $=$ & $K(\lamL{\client}{x}{\ccomp{M}})$
 \\
 & (LamSS) & $\scomp{\lamL{\server}{x}{M}} \ K$ & $=$ & $K(\lamL{\server}{(x,k)}{\scomp{M} \ k})$
 \\ 
 & (AppSC) & $\scomp{\appL{L}{\client}{M}} \ K$ & $=$ & $\scomp{L} \ (\lamL{\server}{f}{}$ \\
  &                &                                                              &         &   $\ \ \ \scomp{M} \ (\lamL{\server}{x}{}$ \\
  &                &                                                              &         &  
              $\ \ \ \ \ \ \call{\ \lamL{\client}{x}{\llet{y}{f(x)}{ \req{K}{y} }}}{\ \ \ x \ } \ ))$            
 \\
 & (AppSS) & $\scomp{\appL{L}{\server}{M}} \ K$ & $=$ & $\scomp{L} \ (\lamL{\server}{f}{}$ \\
  &                &                                                              &         &   $\ \ \ \scomp{M} \ (\lamL{\server}{x}{}$ \\
  &                &                                                              &         &   $\ \ \ \ \ \ f(x,K)))$ 
\end{tabular}
\caption{A typed compilation for \stateencrpc}
\label{fig:compforstateencrpc}
\end{figure*}

	The basic idea of compilation for the server is CPS conversion. The compilation rules for server terms are described as
$
	\scomp{M_{rpc}} \ K = M_{rpc}^{enc}
$,
where $K$ is (a value for) a continuation. In the compiled term, we first evaluate $M_{rpc}$ and then continue with $K$ and the evaluation result. 

	(VarS) compiles a variable into a term applying a continuation $K$ to a value bound to the variable. (LamSC) and (LamSS) have a similar structure as (VarS). They only replace the compiled variable $x$ in (VarS) with $\ccomp{\lamL{a}{x}{M}}$. (AppSS) compiles a local server function application typed under (T-App), and it is the same as the CPS conversion of an application. 
	(AppSC) is the most important compilation rule. A typing derivation denoted by $\appL{L}{\client}{M}$ typed under (T-Call) is compiled as this. Both $L$ and $M$ are first compiled into terms whose evaluation will be bound to a function $f$ and an argument $x$ in the same way as in (AppSS). Then the compilation rule generates 
$
\call{\lamL{\client}{x}{\llet{y}{f(x)}{ \req{K}{y} }} }{x}
$. By this remote client function application,
control is first transferred to the client to evaluate $f(x)$. Control is then returned to the server by $\req{K}{y}$ to continue $K$ in the server with the application result $y$. Our compiler thus supports the behaviour of commuting from  the server to the client by composing this client function, which we call a commute function. 

	In regard to a client function call from the server, the RPC calculus can allow it in non-tail positions but the state-encoding RPC calculus allows it only in the tail position. This mismatch is resolved as follows. 
	The commute function above is composed to have a request with a continuation after the client function call. The continuation will hold the context of the client function call in a non-tail position. 
	In this way, (AppSC) ensures that $\call{-}{-}$ can be always in a tail position in the state-encoding RPC calculus no matter where client function calls from the server appear in the RPC calculus.
	
	Now, we prove the correctness of our compilation for {\stateencrpc} as follows:

\begin{theorem}[Correctness of Compilation for \stateencrpc] Assume a well-typed {\rpc} term $M$ under the locative type system:
\begin{itemize}
\item If $\evalRPCC{M}{V}$, then \runStateEncCSStar{\ccomp{M} \ | \ \emp}{\ccomp{V} \ | \ \emp}.
\item If $\evalRPCS{M}{V}$, then \runStateEncCSStar{\Pi \ | \ \scomp{M} \ K}{\Pi \ | \ \scomp{V} \ K}  for all $\Pi$ and $K$. 
\end{itemize}
\end{theorem}

	The basic idea of our proof is this. A sequence of evaluation steps is built in {\stateencrpc} matched to each subtree of evaluation in {\rpc} by induction on the height of the evaluation trees, and then the sequences are put together to build what is matched to the whole evaluation tree in the conditions. As a natural consequence, the proof also guarantees that client-server communication in {\rpc} is preserved in {\stateencrpc}, though the theorem itself does not state this explicitly. Figure \ref{fig:rpcVSstateless} shows an example of the correspondence of client-server communication between {\rpc} and {\stateencrpc} with identical box names S1, C1, and S2 labelled on the related flows.

\subsection{Separating Client and Server Terms in {\stateenccs}}
\label{sec:stateenccloconv}

Although the state-encoding RPC calculus has the potential to make the {\rpc} terms run on the client and the server separately, in the {\stateencrpc} terms, the client part and the server part are still together. In this section, we propose a state-encoding CS calculus {\stateenccs} where the two parts are clearly separated, and names are used in one part to refer to functions in the other part. 

	Figure \ref{fig:stateenccloconv} shows the syntax and semantics of the state-encoding CS calculus. Terms in the calculus are now denoted by $m$, and values are variables or closures $\clo{F}{\bar{v}}$ denoted by $v$ or $w$. 
	$F$ is a name for a closed function. A closed function, $\bar{z}\lamL{a}{\bar{x}}{m}$, is a function with a list of free variables $\bar{z}$. In a closure, $\bar{v}$ consists of values to substitute for the free variables $\bar{z}$ in the closed function. A substitution of $\bar{v}$ for $\bar{x}$ in $m$ is written as $m\{\bar{v}/\bar{x}\}$.
	The function store $\phi_a$ is a mapping of names into closed functions for each location $a$.
	$\pi$ and $\delta$ are client and server contexts, respectively.
	
	Configurations in the semantics of the state-encoding CS calculus are as follows:
\begin{itemize}
\item $m \ | \ \emp$ for the client to evaluate a term $m$ with the empty server context $\emp$
\item $\pi \ | \ m$ for the server to evaluate a term $m$ with a pending client context $\pi$
\end{itemize}
The semantic rules for the state-encoding CS calculus are isomorphic to those for the state-encoding RPC calculus under the aforementioned changes. 

\begin{figure*}[ht]
\begin{tabular}{l}
\begin{tabular}{l l}
\multicolumn{2}{l}{\textbf{Syntax}} \\
&
\begin{tabular}{l l l c l }
  & Term & $m$    & $::=$  & 
  $v \ | \ v_f(\overline{w}) \ | \ \req{v_f}{\overline{w}} \ | \ \call{v_f}{\overline{w}} \ | \ \llet{x}{m}{m}$  \\
  & Value & $v,w$  & $::=$  & $x \ \ | \ \ \clo{F}{\bar{v}}$ \\ 
  & Function store & $\phi_a$ & $::=$ & $\{ \ \cdots, \ F= \bar{z}\lamL{a}{\bar{x}}{m}, \cdots \ \}$ \\
  & Client context & $\pi$  & $::=$  & $\LetK{x}{m}$               \\
  & Server context & $\delta$  & $::=$  & $\emp$                   
\end{tabular}
\end{tabular}
\\   
\begin{tabular}{l l}
\multicolumn{2}{l}{\textbf{Semantics}} \\
&
\begin{tabular}{l l l  l l l}
\multicolumn{6}{l}{ 
     \framebox{
	\begin{minipage}{0.96\textwidth}
		Client: \ \ \  \runStateEncCS{m \ | \ \emp}{m' \ | \ \emp} \ \ \ or \ \ \ 
     				\runStateEncCS{m \ | \ \emp}{\pi \ | \ m'}
	\end{minipage} 	}
} 
\\[0.2cm]
 \ \ \ (AppC)
 &
 \multicolumn{5}{l}{
 $\llet{y}{\clo{F}{\bar{v}}(\overline{w})}{m} \ | \ \emp
 \ \ \Rightarrow^{enc} \ \ 
 \llet{y}{m_0\substzsxs}{m} \ | \ \emp$
 \ \ if $\phi_\client(F)=\bar{z}\lamL{\client}{\bar{x}}{m_0}$
 }
\\[0.1cm]
 \ \ \ (Req)
 & 
 \multicolumn{5}{l}{
 $\llet{x}{\req{\clo{F}{\bar{v}}}{\overline{w}}}{m} \ | \ \emp
 \ \ \Rightarrow^{enc} \ \ 
 \LetK{x}{m} \ | \ \clo{F}{\bar{v}}(\overline{w})$
 \ \ \ if $\phi_\server(F)=\bar{z}\lamL{\server}{\bar{x}}{m_0}$
 }
\\[0.1cm]
 \ \ \ (ValC)
 & 
 \multicolumn{5}{l}{
 $\llet{x}{v}{m} \ | \ \emp
 \ \ \Rightarrow^{enc} \ \ 
 m\subst{v}{x} \ | \ \emp$}
 \\[0.1cm]
 \ \ \ (LetC)
 & 
 \multicolumn{5}{l}{
 $\llet{x}{  (\llet{y}{m_1}{m_2})  }{m} \ | \ \emp
 \ \ \Rightarrow^{enc} \ \ 
 \llet{y}{m_1}{(\llet{x}{m_2}{m})} \ | \ \emp$}
 \\[0.1cm] 
\end{tabular}
\\
&
\begin{tabular}{l l l l l l}
\multicolumn{6}{l}{ 
     \framebox{
	\begin{minipage}{0.96\textwidth}
			Server: \ \ \ \runStateEncCS{\pi \ | \ m}{\pi \ | \ m'} \ \ \ or \ \ \  
     					\runStateEncCS{\pi \ | \ m}{m' \ | \ \emp}
	\end{minipage}  }
} 
\\[0.2cm]
 \ \ \ (AppS)
 &  
 \multicolumn{5}{l}{
 $\pi \ | \ \clo{F}{\bar{v}}(\overline{w})
 \ \ \Rightarrow^{enc} \ \ 
 \pi \ | \ m_0\substzsxs$ 
 \ \ \ if $\phi_\server(F)=\bar{z}\lamL{\server}{\bar{x}}{m_0}$  }
 \\[0.1cm] 
 \ \ \ (Call)
 &  
 \multicolumn{5}{l}{
 $\LetK{x}{m} \ | \ \call{\clo{F}{\bar{v}}}{\overline{w}} 
 \ \ 
 \Rightarrow^{enc} \ \ 
 \llet{x}{\clo{F}{\bar{v}}(\overline{w})}{m} \ | \ \emp$
 \ \ \ if $\phi_\client(F)=\bar{z}\lamL{\client}{\bar{x}}{m_0}$}
 \\[0.1cm]
 \ \ \ (Reply)
 &  
 \multicolumn{5}{l}{
 $\LetK{x}{m} \ | \ v
 \ \ \Rightarrow^{enc} \ \ 
 \llet{x}{v}{m} \ | \ emp$}
\end{tabular}
\\[0.1cm]
\multicolumn{2}{l}{\textbf{Compilation}} 
\\
&
\begin{tabular}{l l c l l l }
\multicolumn{6}{l}{
  \framebox{$\cconv{M_{rpc}^{enc}}\ = \ m$}
}
\\
 & $\cconv{x} $ & $=$ & $x$ \\
 & $\cconv{\lamL{a}{x}{m}} $ & $=$ & $clo(F,\bar{z})$ \ \ \ if $\bar{z}=fv(\lamL{a}{x}{m})$ and $\phi_a(F)=\bar{z}\lamL{a}{x}{\cconv{m}}$\\
 & $\cconv{v(\overline{w})} $ & $=$ & $\cconv{v}(\ \overline{\cconv{w}} \ )$ \\
 & $\cconv{\llet{x}{m_1}{m_2}}$ & $=$ & $\llet{x}{\cconv{m_1}}{\cconv{m_2}}$ \\
 & $\cconv{\req{v}{\overline{w}}} $ & $=$ & $\req{\cconv{v}}{\ \overline{\cconv{w}} \ }$ \\
 & $\cconv{\call{v}{\overline{w}}} $ & $=$ & $\call{\cconv{v}}{\ \overline{\cconv{w}} \ }$
\end{tabular}
\end{tabular}
\end{tabular}
\\[0.2cm]
\caption{Syntax and semantics of \stateenccs with a compilation of {\stateencrpc}}
\label{fig:stateenccloconv}
\end{figure*}

	Figure \ref{fig:stateenccloconv} also shows the compilation rules of the state-encoding RPC calculus into the state-encoding CS calculus. The compilation rules are actually closure conversion gathering closed functions for each location. For a given {\stateencrpc} term $M$,  the following compilation generates a {\stateenccs} term $m$: 
\[
	\cconv{M} \ = \ m
\]
	The compilation rules traverse a source term. Whenever a location-annotated function is encountered, a fresh name $F$ is created, a list of free variables $\bar{z}$ is collected over the function, and a mapping of this name into a constructed closed function is added to the function store at the location. $fv$ is the standard function for free variable collection trivially extended with the located lambda calculus. Note that the compilation rules utilise implicit global function stores, with which the mapping of names and closed functions encountered during the compilation is checked. Actually, rather than checking the presence of closed functions in the function stores, we intend to collect the closed functions to construct the function stores. 

	Our main compilation method using $\cconv{-}$ compiles a closed state-encoding RPC term into a closed state-encoding CS term. As a side effect, it generates a function store $\phi_\client$ for the client and a function store $\phi_\server$ for the server. Now we can prove that a combined compilation of $\ccomp{-}$ with $\cconv{-}$ of RPC terms will generate state-encoding CS terms that correctly implement the RPC terms as follows.

\begin{theorem}[Correctness of Compilation for \stateenccs] Assume a well-typed term $M$ under the locative type system. Given $\phi_\client$ and $\phi_\server$: 
\begin{itemize}
\item If $\evalRPCC{M}{V}$, then 
\runStateEncCSStar{\cconv{\ccomp{M}} \ | \ \emp}{\cconv{\ccomp{V}} \ | \ \emp}.
\end{itemize}
\end{theorem}

	By this compilation correctness theorem, it is demonstrated that the feature of the RPC calculus allowing arbitrarily deep nesting of control contexts between the client and the server can be correctly implemented by a series of flat request-responses under the stateless server strategy.
	
	We have implemented a parser of {\rpc}, a location-type inference procedure for our type system, a compiler, and an evaluator of {\stateencrpc} and {\stateenccs} on HTTP between the client and the server.  Figure \ref{fig:stateenccsCompExample} shows a compilation of the RPC term in Figure \ref{fig:rpcexample}, generated by the compiler. The evaluator starts with $main$ in the client. For readability, each list of free variables is enclosed by $\{ \ \}$, and so is each list of values for such free variables in closures.
	
\begin{figure}[t]
\begin{tabular}{r c l}	
$\phi_\client$ & : & 
  $main$ \ = \ $\llet{r_3}{\req{clo(g_7, \{\})}{clo(g_{10}, \{\}),clo(g_{11}, \{\})}}{r_3}$ \\
 & &
	$g_2$ = $\{f_7,f_5,k_4\} \ \lamL{\client}{z_9}{\ \llet{r_{10}}{f_7 \ z_9}{\req{clo(g_1,\{f_5,k_4\}}{\  r_{10}}}}$ \\
 & & $g_{10}$ = $\{\} \ \lamL{\client}{y}{\ 
\llet{r_{14}}{\req{clo(g_8, \{\})}{\ y,clo(g_9, \{\})}}{r_{14}}}$ \\
$\phi_\server$ & : & 
 $g_1$ = $\{f_5,k_4\} \ \lamL{\server}{x_6}{\ f_5 \ (x_6,k_4)}$ \\
 & & $g_3$ = $\{f_7,f_5,k_4\} \ \lamL{\server}{x_8}{\ \call{clo(g_2, \{f_7,f_5,k_4\})}{\ x_8}}$ \\
 & & $g_4$ = $\{f_5,k_4\} \ \lamL{\server}{f_7}{\ clo(g_3, \{f_7,f_5,k_4\}) \ c}$ \\
 & & $g_5$ = $\{k_4,f\} \ \lamL{\server}{f_5}{\ clo(g_4, \{f_5,k_4\})) \ f}$ \\
 & & $g_6$ = $\{\} \ \lamL{\server}{x,k_{11}}{ \ k_{11} \ x}$ \\
 & & $g_7$ = $\{\} \ \lamL{\server}{f,k_4}{\ clo(g_5, \{k_4,f\}) \ (clo(g_6, \{\}))}$ \\
 & & $g_8$ = $\{\} \ \lamL{\server}{z,k_{15}}{\ k_{15} \ z}$ \\
 & & $g_9$ = $\{\} \ \lamL{\server}{x_{16}}{\ x_{16}}$ \\
 & & $g_{11}$ = $\{ \} \ \lamL{\server}{x_{17}}{\ x_{17}}$ 
\end{tabular}

\caption{Compilation of the RPC term in Figure \ref{fig:rpcexample}}
\label{fig:stateenccsCompExample}
\end{figure}

	Our state-encoding calculi are higher-order, whereas the original CS calculus remains first-order. This is because the original CS calculus employed an implementation method called {\it defunctionalisation} \cite{Reynolds:1972:DIH:800194.805852}, which converts higher-order functions into first-order ones. However, the state-encoding calculi offer more freedom in the implementation of higher-order functions. They do not actually depend on any specific implementation methods such as closure conversion; they could even utilise defunctionalisation as well.

\subsection{Discussion}
\label{sec:discussionstateencrpc}

\subsubsection{Trampolined Style}

	Our state-encoding calculi use a new trampolined style embedded in the semantics. This contrasts with the original CS calculus \cite{Cooper:2009:RC:1599410.1599439}, which has the trampolined style \cite{Ganz:1999:TS:317636.317779} formulated using terms that introduces a special loop function called {\it trampoline} in the client as follows:
\begin{center}
\begin{tabular}{l l}
Line 1: & $trampoline(x) \ = \ case \ x \ of$ \\
Line 2: & $ \ \ \ \ \ | \ \  Call(f,x,k) \rightarrow trampoline(req(k,f(x)))$ \\
Line 3: & $ \ \ \ \ \ | \ \ Return(x) \rightarrow x$ \\
Line 4: & $trampoline(req(g,v))$ 
\end{tabular}
\end{center}	
	In Line 4, an invocation of a server function $g$ from the client is surrounded by the trampoline function. It is like the semantic behaviour defined by (Req) in the state-encoding calculi. 	
	On receiving a datum like $Call(f,x,k)$ from the server in Line 2, the client applies $f$ to $x$, and then invokes a continuation $k$ with the application result. This is surrounded by the trampoline function again for repetition as many times as is necessary.	
	For commuting between the server and the client like this, our compilation method composes commute functions, as explained in Section \ref{sec:compforstateenccs}:
\[
commute \triangleq 
	\lamL{\client}
			{(f,x,k)}
			{ \llet{y}{ f(x) }{\req{k}{y}} }
\]
so that $Call(f,x,k)$ can be regarded as $commute(f,x,k)$. In the trampoline function, two branches in Lines 2 and 3, one for $Call$ and the other for $Return$, correspond to what are supported by (Call) and (Reply), respectively. 

\subsubsection{A Location-Typed Approach}
	A difference between our state-encoding calculi and the original calculi \cite{Cooper:2009:RC:1599410.1599439} is the use of locative type information for compiling function applications. 
	In the original location-untyped approach, location information is not available for function applications, whereas functions have location annotations. Both the client function and the server function can therefore be invoked in the same function application. For example, in the {\rpc} term:
\[
\llet{g}{\ \lamL{a}{f}{f \ M} \ }
	{\ \cdots \ g \ (\lamL{\client}{x}{x}) \ 
	 \cdots \ g \ (\lamL{\server}{y}{y}) \ 
	 \cdots }	
\]
both $(\lamL{\client}{x}{x})$ and $(\lamL{\server}{y}{y})$ are invoked in the same application term $f \ M$.
	As a result, every function application must check at runtime the location of the function to decide whether to use the trampoline function in case of the server function. In our location-typed approach, local function applications are clearly separated from remote function applications, and so no such dynamic check at runtime is required. We also showed that all terms in the location-untyped approach can be supported in the location-typed approach by the transformation explained in Section \ref{sec:rpc}.
	
	In addition to the absence of runtime checks on location, our location-typed approach results in a  simple form of compilation where there are only two kinds of compilation rules, one for the client and the other for the server. The original compilation method in the location-untyped approach consists of eight kinds of compilation rules: three for the client and five for the server. 
	The simple form helps explain the essential structure of our compilation rules: the compilation of {\rpc} into {\stateencrpc} is a combination of direct-style compilation for the client and CPS-style compilation for the server, and the compilation of {\stateencrpc} into {\stateenccs} is a variant of closure conversion classifying closed functions on locations. 
	The simple compilation rules using locative types facilitate the development of new stateful  calculi in the next section, which was not achieved by the previous work.

\section{Locative Calculi with Explicit Server States}
\label{sec:statefulcalculi}

	 
	The stateless server strategy in the state-encoding calculi is not always satisfactory. When the server states involve disks or databases that do not permit serialisation, it is difficult to encode them.
	Let us see an example:
\[
\mbox{
\begin{tabular}{l l}
$\lamL{\server}{query}{\!\!\!\!\!}$ &
 $\llet{cursor}{ executeOnDatabase(query) }{}$ \\
                                     &
 $\llet{name}{getNameFromRecord(cursor)}{}$ \\
                                     &
 $\llet{r}{f_{client}(name)}{}$ \\
                                     &
 $\llet{cursor}{nextRecord(cursor)}{\ \ \cdots \ \ }$
\end{tabular}
}
\]
	This is a server function that has a query as an argument. It is assumed to invoke a client function repeatedly for each of a list of records obtained from a database query. 
A remote client function invocation $f_{client}(name)$ resides between two subsequent database operations. We should serialise a list of records referenced by the database cursor before the invocation. However, in general, it is difficult to pack the records into a continuation leaving nothing in the server. Even when it is possible to serialise such server states, moving the serialised records between the client and the server repeatedly will increase communication overheads. This will cause some efficiency concerns. This observation motivates us to design new stateful calculi that eliminate the need to encode server states.

	In this section, we develop new calculi that save and restore server states using the stack in the underlying semantics, and therefore there is no need to encode them as in the state-encoding calculi.  In fact, using the runtime stack is similar to how conventional programming languages have approached the problem of saving and restoring control contexts. 

	We first introduce the syntax and semantics for a stateful RPC calculus {\statefulrpc} using stack-based control states in the server. Then we propose a typed compilation method of compiling {\rpc} terms into {\statefulrpc} terms and subsequently another compilation method for separating client and server terms in {\statefulcs}. 

\subsection{A Stateful RPC Calculus \statefulrpc}
\label{sec:statefulcs}

Let us explore the features of a stateful RPC calculus {\statefulrpc} before we explain the details of its syntax and semantics. An important feature is a stack $\Delta$ in the server:
\[
\Delta\ ::= \ \emp \ \ | \ \ \LetK{x}{M}\cdot\Delta
\]
where the empty stack is denoted by $\emp$, and any non-empty stack consists of contexts concatenated with the dot operator. A context $\LetK{x}{M}$ can be viewed as a let term with a hole such as $\llet{x}{[-]}{M}$ denoting a control state. 

	We illustrate two new constructs $\textsfCall$ and $\textsfRet$ in the stateful RPC calculus with the following steps under the evaluation relation $\Rightarrow^{state}$:
	\begin{center}
	\begin{tabular}{l l l l l}
	 & \ \ \ \ [Client]                   &               & [Server] \\ 
	 & $\ \ \ \  \LetK{z}{M_z}$ & $ \ \ | \ \ $ & $ \Delta; \ \llet{x}{\call{f_{client}}{V_{arg}}}{M}$ & (1) \\
	 $\Rightarrow^{state}$ & $ \ \ \ \ \llet{z}{f_{client}(V_{arg})}{M_z}$ & $  \ \ | \ \ $ & $ \LetK{x}{M}\cdot\Delta$ & (2) \\
	 $\Rightarrow^{state \ +}$ & $ \ \ \ \ \llet{z}{\ret{W}}{M_z}$ & $ \ \ | \ \ $ & $ \LetK{x}{M}\cdot\Delta$ & (3) \\
	 $\Rightarrow^{state}$ & $ \ \ \ \ \LetK{z}{M_z}$ & $ \ \ | \ \ $ & $ \Delta; \ \llet{x}{W}{M}$ & (4) 
	\end{tabular} 
	\end{center}
	
	In (1), a server is about to switch to a client in order to invoke a client function. The new $\textsfCall$ construct does this after pushing the server context ($\LetK{x}{M}$) onto a stack $\Delta$. 
	In (2), the waiting client context ($\LetK{z}{M_z}$) is open to become the let with a hole filled with the client function application $f_{client}(V_{arg})$, and the server context stack grows from $\Delta$ to $\LetK{x}{M}\cdot\Delta$. 
	This application is assumed to evaluate eventually to $\ret{W}$ using the other new construct. 
	In (3), the new $\textsfRet$ construct is about to switch to the server to return the value $W$ after popping the top server context from the server stack and restoring the server context saved by $\textsfCall$ previously. In addition, the $\textsfRet$ construct leaves the client context surrounding itself in the client. This will lead to (4). 

	Note that $\call{f}{\overline{W}}$ in {\statefulrpc} is different from $\call{f}{\overline{W}}$ in {\stateencrpc} in regard to whether or not the server stack is involved. Accordingly, the organisation of client functions for calling by the constructs is different, as we will see soon.

	The configuration `$\llet{x}{\ret{V}}{M} \ | \ \Delta$' in the stateful RPC calculus is reminiscent of the configuration `$\llet{x}{\req{K}{V}}{M} \ | \ \emp$' in the state-encoding RPC calculus. The server stack $\Delta$ in the former can be regarded as the continuation $K$ encoded for a server control state in the latter. As we apply $K$ to $V$ in the server and bind the application result to $x$ in the latter, we return $V$ to the server with $\Delta$ and bind the evaluation result to $x$ in the former.  

	Consider again the {\rpc} example term:
\[
(\lamL{\server}{f}{ \ (\lamL{\server}{x}{x}) \ (f \ \ c) }) \ \ (\lamL{\client}{y}{  \ (\lamL{\server}{z}{z}) \ \ y   })
\]
For comparison with the state-encoding RPC calculus, this term can be implemented in the stateful RPC calculus as
\[
\llet{r}{\req{M_1}{M_2}}{r}
\]
where $M_1$ and $M_2$ are supposed to implement the outermost function $(\lamL{\server}{f}{ \ (\lamL{\server}{x}{x}) \ (f \ \ c) })$ and its argument $(\lamL{\client}{y}{  \ (\lamL{\server}{z}{z}) \ \ y   })$, respectively.
	The use of {\textsfReq} is the same as for {\stateencrpc}. The notable difference is that it does not require any continuation argument. In {\statefulrpc}, it is not necessary to use CPS conversion for server functions, because the underlying semantics of the stateful calculus explicitly manages a server stack that continuations would substitute for in the state-encoding calculus. 

	Now let us examine $M_1$:
\[ 
\mbox{
\begin{tabular}{l}
$M_1 \triangleq \lamL{\server}{f}{ \ }
 \ \llet{w}{\call{\ \ \lamL{\client}{x}{\llet{y}{f(x)}{\ret{y}}} \ \ }{\ \ [\![ c  ]\!]}}{}$ \\                      
$\ \ \ \ \ \ \ \ \ \ \ \  \ \ \ \ \  \ \   \llet{r}{[\![ \lamL{\server}{x}{x} ]\!](w)}{r}$
\end{tabular}
}
\]
where $[\![M]\!]$ represents an implementation of a {\rpc} term $M$ for the server. Given $f$ as a client function, {\textsfCall} switches to the client to invoke a commute function
$(\lamL{\client}{x}{\llet{y}{f(x)}{\ret{y}}})$,
pushing a server context 
$(\LetK{w}{(\llet{r}{[\![ \lamL{\server}{x}{x} ]\!](w)}{r})})$ 
on a server stack.
	The commute function applies $f$ to $x$ (i.e. $[\![ c  ]\!]$) in the client. It then  returns the result value $y$ to the server by $\ret{y}$, popping the server context. We continue to evaluate the term from the restored context with the result value bound to $w$ in the server.
		
	One implementation of $M_2$ is as follows:
\[
M_2 \triangleq\lamL{\client}{y}{ \ 
 \llet{ r}{\req{[\![\lamL{\server}{z}{z}]\!]}{ \ y} }{r} 
 }
\]
When compared with the implementation for {\stateencrpc}, this is a simple structure with no continuation argument in the use of {\textsfReq} to invoke a server function. 

\begin{figure*}
\begin{tabular}{l}
\begin{tabular}{l l}
\multicolumn{2}{l}{\textbf{Syntax}} \\
&
\begin{tabular}{l l l c l }
   & Term & $M$    & $::=$  & $V \ | \ V_f(\overline{W}) \ | \ 
   		\req{V_f}{\overline{W}} \ | \
   		\call{V_f}{\overline{W}} \ | \
   		\ret{V} \ | \ \llet{x}{M}{M} $                  \\
  & Value & $V,W$  & $::=$  & $x \ \ | \ \ \lamL{a}{\bar{x}}{M}$\\ 
  & Client context & $\Pi$  & $::=$ & $\LetK{x}{M}$ \\
  & Server context stack & $\Delta$  & $::=$ & $\emp \ \ | \ \ \LetK{x}{M}\cdot\Delta$ 
\end{tabular}
\end{tabular}
\\
\begin{tabular}{l l}
\multicolumn{2}{l}{\textbf{Semantics}} \\
&
\begin{tabular}{l l l l l l }
\multicolumn{6}{l}{ 
     \framebox{
	\begin{minipage}{0.94\textwidth}
			Client:  \ \ \ \runStatefulCS{M \ | \ \Delta}{M' \ | \ \Delta} \ \ \ or \ \ \   
     						\runStatefulCS{M \ | \ \Delta}{\Pi \ | \ \Delta'; M'}
	\end{minipage} }
} 
\\[0.2cm]
 \ \ \  (AppC) 
 & 
 \multicolumn{5}{l}{
 $\llet{y}{(\lamL{\client}{\bar{x}}{M_0})(\overline{W})}{M} \ | \ \Delta
 \ \ \Rightarrow^{state} \ \ 
 \llet{y}{M_0\substXs}{M} \ | \ \Delta$}
 \\[0.1cm]
 \ \ \ (Req)
 &
 \multicolumn{5}{l}{
 $\llet{x}{\req{\lamL{\server}{\bar{x}}{M_0}}{\overline{W}}}{M} \ | \ \Delta
 \ \ \Rightarrow^{state} \ \ 
 \LetK{x}{M} \ | \ \Delta; \ \llet{r}{(\lamL{\server}{\bar{x}}{M_0})(\overline{W})}{r}$}
 \\[0.1cm] 
 \ \ \  (ValC)
 &
 \multicolumn{5}{l}{
 $\llet{x}{V}{M} \ | \ \Delta
 \ \ \Rightarrow^{state} \ \ 
 M\subst{V}{x} \ | \ \Delta$}
 \\[0.1cm] 
 \ \ \  (LetC)
 &
 \multicolumn{5}{l}{
 $\llet{x}{  (\llet{y}{M_1}{M_2})  }{M} \ | \ \Delta
 \ \ \Rightarrow^{state} \ \ 
 \llet{y}{M_1}{(\llet{x}{M_2}{M})} \ | \ \Delta$}
 \\[0.1cm]
 \ \ \ (Ret) 
 &
 \multicolumn{5}{l}{
 $\llet{y}{\ret{V}}{M_2} \ | \ \LetK{x}{M_1}\cdot\Delta
 \ \ \Rightarrow^{state} \ \ 
 \LetK{y}{M_2} \ | \ \Delta; \ \llet{x}{V}{M_1}$}
 \\[0.1cm]  
\end{tabular}
 \\
&
\begin{tabular}{l l l  l l l }
 \multicolumn{6}{l}{ 
     \framebox{
	\begin{minipage}{0.94\textwidth}
			Server:  \ \ \ \runStatefulCS{\Pi \ | \ \Delta; M}{\Pi \ | \ \Delta; M'} \ \ \ or \ \ \   
     						\runStatefulCS{\Pi \ | \ \Delta; M}{M' \ | \ \Delta'}
	\end{minipage}  }
 } 
 \\[0.2cm]
 \ \ \  (AppS)
 &
 \multicolumn{5}{l}{
 $\Pi \ | \ \Delta; \ \llet{y}{(\lamL{\server}{\bar{x}}{M_0})(\overline{W})}{M} 
 \ \ \Rightarrow^{state} \ \ 
 \Pi \ | \ \Delta; \ \llet{y}{M_0\substXs}{M}$}
 \\[0.1cm] 
 \ \ \ (Call) 
 &
 \multicolumn{5}{l}{
 $\LetK{y}{M_2} \ | \ \Delta; \ \llet{x}{\call{\lamL{\client}{\bar{x}}{M_0}}{\overline{W}}}{M_1}
 $}
 \\[0.1cm]
 & 
 \multicolumn{5}{l}{
 $
 \ \ \Rightarrow^{state} \ \ 
 \llet{y}{(\lamL{\client}{\bar{x}}{M_0})(\overline{W})}{M_2} \ | \ \LetK{x}{M_1}\cdot\Delta $}
 \\[0.1cm]
 \ \ \  (Reply) 
 &
 \multicolumn{5}{l}{
 $\LetK{x}{M} \ | \ \Delta; \ V
 \ \ \Rightarrow^{state} \ \ 
 \llet{x}{V}{M} \ | \ \Delta$}
 \\[0.1cm] 
 \ \ \  (ValS)
 &
 \multicolumn{5}{l}{
 $\Pi \ | \ \Delta; \ \llet{x}{V}{M}
 \ \ \Rightarrow^{state} \ \ 
 \Pi \ | \ \Delta; \ M\subst{V}{x}$}
 \\[0.1cm] 
 \ \ \ (LetS)
 &
 \multicolumn{5}{l}{
 $\Pi \ | \ \Delta; \ \llet{x}{  (\llet{y}{M_1}{M_2})  }{M}
 \ \ \Rightarrow^{state} \ \
 \Pi \ | \ \Delta; \ \llet{y}{M_1}{(\llet{x}{M_2}{M})}$}
\end{tabular}
\end{tabular}
\end{tabular}
\\[0.2cm]
\caption{A stateful client-server calculus \statefulrpc}
\label{fig:statefulrpc}
\end{figure*}

\subsection{The Formal Semantics of \statefulrpc}
\label{sec:synsemstatefulcs}

	Figure \ref{fig:statefulrpc} shows the syntax and semantics of the stateful RPC calculus {\statefulrpc} in the same style of the state-encoding RPC calculus. Basically, the syntax of the calculus has A-normal form with three constructs: $\textsfReq$, $\textsfCall$, and $\textsfRet$. 
	We intend this new construct $\ret{V}$ to return a value to the server from the client, and a symmetric construct of $\call{f}{\overline{W}}$ invokes a client function from the server. 
	
	Values are either a variable or a locative function. Client contexts $\Pi$ are the same for {\stateencrpc}. Server contexts now have the form of a stack denoted by $\Delta$.
		
	There are two kinds of configurations in the stateful RPC calculus: 
	\begin{itemize}
	\item $M \ | \ \Delta$ for the client to evaluate a term $M$ with a stack $\Delta$ in  the server separated by $|$  
	\item $\Pi \ | \ \Delta; M$ for the server to evaluate a term $M$ on a stack $\Delta$  with a pending client context $\Pi$ in the client separated by $|$ 
	\end{itemize}
	
	The semantics of the calculus is described by the evaluation relation $\Rightarrow^{state}$ on the configurations. The semantic rules are defined in a minimal way for {\statefulrpc} enough to be an intermediate calculus for compilation of {\rpc} as those for {\stateencrpc}.
	In the semantics, (AppC), (ValC), and (LetC) are semantic rules for locally running the client in the form \runStatefulCS{M \ | \ \Delta}{M' \ | \ \Delta}. 
	Note that the three client-side rules are different from those rules in {\stateencrpc} only in that the server context stack $\Delta$ replaces the empty server context of the rules in {\stateencrpc}. 
	(AppS), (ValS), and (LetS) are  semantic rules for locally running the server in the form  \runStatefulCS{\Pi \ | \ \Delta;M}{\Pi \ | \ \Delta; M'}. 
	The three server-side rules introduce the server context stack $\Delta$ when compared with those rules in {\stateencrpc}. 
	
	The remaining four semantic rules, (Req), (Ret), (Call), and (Reply), change active running from the client to the server or vice versa. We intend a request initiated by (Req) to finish by (Reply), and we also intend a call that (Call) begins to finish with (Ret). 
	Note that both (Req) and (Reply) are exactly the same as those in {\stateencrpc} except for the introduction of a server context stack $\Delta$ instead of the empty server context. 
	
	(Call) in {\statefulrpc} exhibits an important difference from that in {\stateencrpc}. The call construct can now be used in a non-tail position, e.g. $\llet{y}{[-]}{M_0}$. (Call) pushes it onto the current server context stack $\Delta$ as $\LetK{y}{M_0}\cdot\Delta$, switching to the client. (Ret) reverses this procedure: it returns a value computed in the client back to the server, popping the top server context from the server stack $\LetK{y}{M_0}\cdot\Delta$.
	
	Interestingly, we find that the client-server interactions in {\statefulrpc} are analogous to the behaviour of coroutines \cite{Conway:1963:DST:366663.366704}. When the client is assumed to be an initiator and the server is assumed to be a coroutine, (Req) is an initiation of a coroutine, (Call) is a suspension of the execution of the coroutine to return to its initiator, (Ret) is a resumption of the coroutine, and (Reply) is a terminator of the courotine. 
	
	The stateful RPC calculus is also based on the semantic-based trampoline style, which was explained for the state-encoding calculi in Section \ref{sec:discussionstateencrpc}, by (Req), (Call), and (Reply) together with the support for composing commute functions by its compiler, which will be explained in the next section.

\begin{figure*}[t]
\begin{tabular}{l l l c l }
 \multicolumn{5}{l}{
  \framebox{Client: \ $\ccomp{M_{rpc}} \ = \ M_{rpc}^{state}$}
}
\\
 & (VarC)  & $\ccomp{x}$ & $=$ & $x$
\\
 & (LamCC) & $\ccomp{\lamL{\client}{x}{M}}$ & $=$ & $\lamL{\client}{x}{\ccomp{M}}$
  \\
 & (LamCS) & $\ccomp{\lamL{\server}{x}{M}}$ & $=$ & $\lamL{\server}{x}{\scomp{M}}$
  \\
 & (AppCC) & $\ccomp{\appL{L}{\client}{M}}$ & $=$ & $\llet{f}{\ccomp{L}}{}$ \\
  &                &                                                          &         &  $\llet{x}{\ccomp{M}}{}$ \\
  &                &                                                          &         &  $\llet{r}{f(x)}{r}$ 
 \\
 & (AppCS) & $\ccomp{\appL{L}{\server}{M}}$ & $=$ & $\llet{f}{\ccomp{L}}{}$ \\
  &                &                                                          &         &  $\llet{x}{\ccomp{M}}{}$ \\
  &                &                                                          &         &  $\llet{r}{\req{f}{x}}{r}$            
\\[0.1cm]
 \multicolumn{5}{l}{
  \framebox{Server: \ $\scomp{M_{rpc}} \ = \ M_{rpc}^{state}$}
}
\\
 & (VarS)  & $\scomp{x}$ & $=$ & $x$
\\
 & (LamSC) & $\scomp{\lamL{\client}{x}{M}}$ & $=$ & $\lamL{\client}{x}{\ccomp{M}}$
 \\
 & (LamSS) & $\scomp{\lamL{\server}{x}{M}}$ & $=$ & $\lamL{\server}{x}{\scomp{M}}$
\\
 & (AppSC) & $\scomp{\appL{L}{\client}{M}}$     & $=$ &  $\llet{f}{\scomp{L}}{}$ \\
  &                &                                                              &         &  $\llet{x}{\scomp{M}}{}$ \\
  &                &                                                              &         &  
              $\llet{r}{\call{\ \lamL{\client}{x}{\llet{y}{f(x)}{ \ret{y} }}}{\ \ \ x \ }}{r}$            
 \\
 & (AppSS) & $\scomp{\appL{L}{\server}{M}}$    & $=$ &   $\llet{f}{\scomp{L}}{}$ \\
  &                &                                                              &         &   $\llet{x}{\scomp{M}}{}$ \\
  &                &                                                              &         &   $\llet{r}{f(x)}{r}$ 
\\
\end{tabular}
\caption{A typed compilation for \statefulrpc}
\label{fig:compforstatefulcs}
\end{figure*}

\subsection{A Typed Compilation of {\rpc} for {\statefulrpc}}
\label{sec:compforstatefulcs}

	Now we present a typed compilation of the RPC calculus into the stateful RPC calculus. Our compilation takes a typing derivation of a source term under our locative type system in the same way as before. 

	Figure \ref{fig:compforstatefulcs} shows our compilation rules. The compilation rules for the client and the server have the same form:   
\[
	\ccomp{M_{rpc}} \ = \ M_{rpc}^{state} \ \ \ \ \mbox{and} \ \ \ \
	\scomp{M_{rpc}} \ = \ M_{rpc}^{state}
\]	
	The compilation rules are in direct style. 
	This contrasts with the use of CPS conversion in the compilation rules for the state-encoding RPC calculus, which does not support saving and restoring server contexts in its semantics. Therefore, it is a compiler that supports this by CPS conversion. 
	The stateful calculus, however, does support saving and restoring server contexts in the underlying semantics, and so the compiler for this calculus does not need to track any server contexts to manage. 
	As a result, it is not necessary to adopt CPS conversion for compilation of server terms anymore. 

	In the compilation rules, (VarC), (LamCC), and (LamCS) are the same as (VarS), (LamSC), and (LamSS) for variable, client function, and server function. The compilation rules only follow the structure of the terms.
	
	(AppCC) and (AppSS) are for local applications in the client and server, respectively. They generate local application terms $f(x)$ with $f$ and $x$ as the compiled terms of the function and the argument of the applications. 
	
	(AppCS) and (AppSC) are the most interesting compilation rules. (AppCS) compiles a term of invocation of  a server function in the client. It simply generates a target term with $\req{f}{x}$ with $f$ as a compiled server function and $x$ as a compiled argument. 
	Note that there is no identity continuation once introduced in compilation for the state-encoding RPC calculus. 
	(AppSC) compiles a term of invocation of a client function in the server, into a target term with
$
\call{\lamL{\client}{x}{\llet{y}{f(x)}{ \ret{y} }}}{x}
$
composing a commute function with {\textsfCall} and {\textsfRet}.

We now prove a slightly stronger version of the correctness of the compilation for {\statefulrpc}. For the proof, we introduce a definition of {\it call-return balanced} evaluation steps. {\textsfCall} and {\textsfRet} are the only constructs to push onto and to pop from a server stack. Intuitively, this implies that in the evaluation steps, every invocation of a client function from the server will return to the server without changing a client context and a server stack.

\begin{definition} A sequence of evaluation steps, \runStatefulCSStar{Conf_1}{Conf_n}, is said to be call-return balanced if the evaluation sequence is derived from the following grammar:
\begin{itemize}
\item $bal \ \ ::= \ \ Conf_{call} \ \ bal \ \ Conf_{ret} 
           \ \ | \ \ Conf_{else}
           \ \ | \ \ bal_1 \ \ bal_2$ where
\begin{itemize}
\item $Conf_{call}$ and $Conf_{ret}$ : 
			configurations paired in the form of 
			\begin{itemize}
				\item[(1)] $\LetK{y}{M_y} \ | \ \Delta; \ \llet{x}{\call{V}{\overline{W}}}{M_x}$ \ and 
				\item[(2)] $\llet{y}{\ret{V_r}}{M_y} \ | \ \LetK{x}{M_x}\cdot\Delta$,  \ respectively.
			\end{itemize}			
\item $Conf_{else}$ : configurations in the form of neither $Conf_{call}$ nor $Conf_{ret}$
\end{itemize}
\end{itemize}
\label{def:statefulrpcbalanced}
\end{definition}

	Note that it is easy to construct \runStatefulCSStar{Conf_{call}}{\LetK{y}{M_y} \ | \ \Delta; \ \llet{x}{V_r}{M_x}} in the call-return balanced evaluation steps, because \runStatefulCS{Conf_{ret}}{\LetK{y}{M_y} \ | \ \Delta; \ \llet{x}{V_r}{M_x}} by the definition of (Ret). The client context $\LetK{y}{M_y}$ and the server stack $\Delta$ are preserved during the client function call. 

\begin{theorem}[Correctness of Compilation for \statefulrpc] Assume a well-typed term {\rpc} $M$ under the locative type system:
\begin{itemize}
\item If $\evalRPC{M}{c}{V}$, then \runStatefulCSStar{\ccomp{M} \ | \ \Delta}{\ccomp{V} \ | \ \Delta} for all $\Delta$, which is call-return balanced.
\item If $\evalRPC{M}{s}{V}$, then 
\runStatefulCSStar{\Pi \ | \ \Delta; \scomp{M}}{\Pi \ | \ \Delta; \scomp{V}} for all $\Pi$ and $\Delta$, which is call-return balanced.
\end{itemize}
\end{theorem}

\begin{figure*}[ht]
\begin{tabular}{l}
\begin{tabular}{l l}
\multicolumn{2}{l}{\textbf{Syntax}} 
\\
&
\begin{tabular}{l l l c l }
  & Term & $m$    & $::=$  & $v \ | \ v_f(\overline{w}) \ | \ \req{v_f}{\overline{w}} \ | \ \call{v_f}{\overline{w}} \ | \ \ret{v} \ | \ \llet{x}{m}{m}$ \\
  & Value & $v,w$  & $::=$  & $x \ \ | \ \ \clo{F}{\bar{v}}$  \\ 
  & Client context & $\pi$  & $::=$  & $\LetK{x}{m}$   \\
  & Server context & $\delta$  & $::=$  & $\emp \ | \ \LetK{x}{m}\cdot\delta$          
\end{tabular}
\end{tabular}   
\\
\begin{tabular}{l l}
\multicolumn{2}{l}{\textbf{Semantics}} 
\\
&
\begin{tabular}{l l l   l l l }
\multicolumn{6}{l}{
     \framebox{
	\begin{minipage}{0.97\textwidth}
			Client:  \ \ \ \runStatefulCS{m \ \ | \ \ \delta}{m' \ \ | \ \ \delta} \ \ \ or \ \ \   
    						 \runStatefulCS{m \ \ | \ \ \delta}{\pi \ \ | \ \ \delta'; \ m'}
	\end{minipage}  }
 }
\\[0.2cm]
\ \ \ (AppC)
 &
 \multicolumn{5}{l}{
 $\llet{y}{v_F(\overline{w})}{m} \ | \ \delta 
 \ \ \Rightarrow^{state} \ \ 
 \llet{y}{m_0\substzsxs}{m} \ | \ \delta$}
\\
 &
 \multicolumn{5}{l}{
 \ \ \ if $v_F=\clo{F}{\bar{v}}$ and $\phi_\client(F)=\bar{z}\lamL{\client}{\bar{x}}{m_0}$}
 \\[0.1cm]
 \ \ \  (Req)
 &
 \multicolumn{5}{l}{
 $\llet{x}{\req{v_F}{\overline{w}}}{m} \ | \ \delta
 \ \ \Rightarrow^{state} \ \ 
 \LetK{x}{m} \ | \ \delta; \ \llet{r}{v_F(\overline{w})}{r}$
 }
 \\
 &
 \multicolumn{5}{l}{
 \ \ \ if $v_F=\clo{F}{\bar{v}}$ and $\phi_\server(F)=\bar{z}\lamL{\server}{\bar{x}}{m_0}$}
 \\[0.1cm]
 \ \ \ (ValC)
 &
 \multicolumn{5}{l}{
 $\llet{x}{v}{m} \ | \ \delta
 \ \ \Rightarrow^{state} \ \ 
 m\subst{v}{x} \ | \ \delta$}
 \\[0.1cm] 
 \ \ \ (LetC)
 &
 \multicolumn{5}{l}{
 $\llet{x}{  (\llet{y}{m_1}{m_2})  }{m} \ | \ \delta
 \ \ \Rightarrow^{state} \ \ 
 \llet{y}{m_1}{(\llet{x}{m_2}{m})} \ | \ \delta$}
 \\[0.1cm]
 \ \ \ (Ret) 
 &
 \multicolumn{5}{l}{
 $\llet{y}{\ret{v}}{m_2} \ | \  \LetK{x}{m_1}\cdot\delta 
 \ \ \Rightarrow^{state} \ \ 
 \LetK{y}{m_2} \ | \ \delta; \ \llet{x}{v}{m_1}$}
 \\[0.1cm]  
\end{tabular}
 \\
&
\begin{tabular}{l l l   l l l }
\multicolumn{6}{l}{
     \framebox{
	\begin{minipage}{0.97\textwidth}
			Server:  \ \ \ \runStatefulCS{\pi \ \ | \ \ \delta; \ m}{\pi \ \ | \ \ \delta; \  m'} \ \ \ or \ \ \   
     						\runStatefulCS{\pi \ \ | \ \ \delta; \ m}{m' \ \ | \ \ \delta'}
	\end{minipage}  }
 }
\\[0.2cm]
 \ \ \ (AppS) 
 & 
 \multicolumn{5}{l}{
 $\pi \ | \ \delta; \ \llet{y}{v_F(\overline{w})}{m} 
 \ \ \Rightarrow^{state} \ \ 
 \pi \ | \ \delta; \ \llet{y}{m_0\substzsxs}{m}$}
\\
 &
 \multicolumn{5}{l}{
 \ \ \  if $v_F=\clo{F}{\bar{v}}$ and $\phi_\server(F)=\bar{z}\lamL{\server}{\bar{x}}{m_0}$ }
 \\[0.1cm] 
 \ \ \ (Call) 
 &
 \multicolumn{5}{l}{
 $\LetK{y}{m_2} \ | \ \delta; \ \llet{x}{\call{v_F}{\overline{w}}}{m_1} 
 \ \ \Rightarrow^{state} \ \ 
 \llet{y}{v_F(\overline{w})}{m_2} \ | \ \LetK{x}{m_1}\cdot\delta$}
\\
 &
 \multicolumn{5}{l}{
 \ \ \ if $v_F=\clo{F}{\bar{v}}$ and $\phi_\client(F)=\bar{z}\lamL{\client}{\bar{x}}{m_0}$}
 \\[0.1cm]
 \ \ \ (Reply) 
 & 
 \multicolumn{5}{l}{
 $\LetK{x}{m} \ | \ \delta; \ v
 \ \ \Rightarrow^{state} \ \ 
 \llet{x}{v}{m} \ \ | \ \ \delta$}
 \\[0.1cm] 
 \ \ \ (ValS) 
 & 
 \multicolumn{5}{l}{
 $\pi \ | \ \delta; \ \llet{x}{v}{m}
 \ \ \Rightarrow^{state} \ \ 
 \pi \ | \ \delta; \ m\subst{v}{x}$}
 \\[0.1cm]   
 \ \ \ (LetS) 
 & 
 \multicolumn{5}{l}{
 $\pi \ | \ \delta; \ \llet{x}{  (\llet{y}{m_1}{m_2})  }{m}
 \ \ \Rightarrow^{state} \ \ 
 \pi \ | \ \delta; \ \llet{y}{m_1}{(\llet{x}{m_2}{m})}$}
\end{tabular}
\\
\end{tabular}
\\
\begin{tabular}{l l}
\multicolumn{2}{l}{\textbf{Compilation} } 
\\
&
\begin{tabular}{l l c l l l }
\multicolumn{6}{l}{
  \framebox{$\cconv{M_{rpc}^{state}} \ = \ m$}
}
\\
& $\cconv{\ret{v}} $ & $=$ & $\ret{\cconv{v}}$ 
  \ \ \ \ \ (The rest of the compilation rules are the same as for {\stateenccs})
\end{tabular}
\end{tabular}
\end{tabular}
\caption{Syntax and semantics of \statefulcs with a compilation of {\statefulrpc}}
\label{fig:statefulcs}
\end{figure*}

	The basic idea of our proof has the same structure as the proof for {\stateencrpc} explained in Section \ref{sec:compforstateenccs}. The proof builds a sequence of evaluation steps in {\statefulrpc} matched to each subtree of evaluation in {\rpc} by induction on the height of evaluation trees, and then it puts the sequences together to build what is matched to the whole evaluation tree in the conditions. The proof therefore guarantees that client-server communication in {\rpc} is preserved in {\stateencrpc}, though the theorem itself does not tell this explicitly. Later, we will present an example of a correspondence of client-server communication between {\rpc} and {\statefulrpc} indirectly by comparing related flows between {\stateencrpc} and {\statefulrpc} in Section \ref{subsec:sessionmngt}. 

\subsection{Separating Client and Server Terms in {\statefulcs}}
\label{sec:statefulcloconv}

As was done for the state-encoding calculi, we propose a stateful CS calculus.
Again, the basic idea is to use closure conversion to decompose a term into closed functions and to classify them according to location annotations. 

	In Figure \ref{fig:statefulcs}, the syntax and semantics of {\statefulcs} are shown. The syntax of {\statefulcs} is identical to that of {\stateenccs} except for $\ret{v}$. The terms in the calculus are denoted by $m$. The calculus explicitly deals with closures $v$ as values, rather than lambda terms. $F$ is a name for a closed function. $\phi_a$ is a mapping of names into closed functions in the location $a$. $\pi$ and $\delta$ are client and server contexts, respectively.
	
	Configurations in the semantics of {\statefulcs} are changed accordingly:
\begin{itemize}
\item $m \ | \ \delta$ for the client to evaluate a term $m$ with a stack $\delta$ in the server separated by $|$
\item $\pi \ | \ \delta;m$ for the server to evaluate a term $m$ on a stack $\delta$ with a pending client context $\pi$ in the client separated by $|$ 
\end{itemize} 
	The semantic rules for the stateful CS calculus are isomorphic to those for the stateful RPC calculus under the aforementioned changes. 

	The compilation rules for {\statefulcs} are the same as those for {\stateenccs} except for $\ret{v}$ in Figure \ref{fig:statefulcs}. 
	For a given closed {\statefulrpc} term $M$ , $\cconv{M} = m$ means that it generates a {\statefulcs} term $m$ for the client, producing a function store $\phi_\client$ for the client and another function store $\phi_\server$ for the server.

	We now prove the correctness of the compilation for {\statefulcs} by assuming a similar definition of call-return balance for evaluation steps as Definition \ref{def:statefulrpcbalanced}.

\begin{theorem}[Correctness of Compilation for \statefulcs] Assume a well-typed term $M$ under the locative type system. Given $\phi_\client$ and $\phi_\server$: 
\begin{itemize}
\item If $\evalRPCC{M}{V}$, then \\ 
\runStatefulCSStar{\cconv{\ccomp{M}} \ | \ \emp}{\cconv{\ccomp{V}} \ | \ \emp}, which is call-return balanced.
\end{itemize}
\end{theorem}

	While we reused the parser of {\rpc}  and the location-type inference procedure for our type system
mentioned in Section \ref{sec:stateenccloconv}, we have implemented a compiler and an evaluator of {\statefulrpc} and {\statefulcs} on HTTP between the client and the server. 
	Figure \ref{fig:statefulcsCompExample} shows a compilation of the RPC term in Figure \ref{fig:rpcexample}, generated by the compiler. The evaluator starts with $main$ in the client.
	
\begin{figure}[t]
\begin{tabular}{r c l}	
$\phi_\client$ & : & 
$main$ \ = \ $\llet{r_3}{\req{clo(g_3, \{\})}{\ clo(g_5, \{\})}}{r_3}$ \\
 & & 
 $g_2$ = $\{f_7\} \ \lamL{\client}{z_{10}}{\ \llet{y_9}{f_7 \ z_{10}}{\ret{y_9}}}$ \\
 & & $g_5$ = $\{\} \ \lamL{\client}{y}{\ \llet{r_{14}}{\req{clo(g_4, \{\})}{\ y}}{r_{14}}}$ \\
$\phi_\server$ & : & 
 $g_1$ = $\{\} \ \lamL{\server}{x}{ \ x}$ \\
 & & $g_3$ = $\{\} \ \lamL{\server}{f}{\ \llet{x_5}{(\llet{r_{11}}{\call{clo(g_2, \{f\})}{\ c}}{ r_{11}})}{\llet{r_6}{clo(g_1, \{\}) \ x_5}{r_6}}}$ \\
 & & $g_4$ = $\{\} \ \lamL{\server}{z}{\ z}$ 
\end{tabular}
\caption{A compilation of the RPC term in Figure \ref{fig:rpcexample}}
\label{fig:statefulcsCompExample}
\end{figure}

\subsection{Discussion}
\label{sec:discussionstatefulrpc}

\subsubsection{Session Management}
\label{subsec:sessionmngt}
	Because the RPC calculus is designed for the web-based client-server model, it is necessary to discuss whether or not the new stateful CS calculus fits the model. 
	In the state-encoding CS calculus, it is rather straightforward to associate the semantics with the {\it request-response} interaction of the web-based model, thanks to the nature of the {\it stateless} server which is aligned well with the RESTful web. 
	However, in the stateful CS calculus, it is not so obvious how the semantics is connected with the web-based interaction, because the web-based model does not directly support the server states that the semantics deals with.  
	
	To explain how the stateful CS calculus manages server states through a series of multiple request-response interactions, we introduce the notion of {\it session} to capture one or more request-response interactions. In the notation, we place a session annotation over the bar separating a client and a server as
	\begin{itemize}
	\item `$client \ \overset{\sessionSomething}{|} \ server$' to denote the client and the server connected under a session identified by a unique number $session$, or 
	\item `$client \ \overset{\sessionNothing}{|} \ server$' to denote that no session is established between the two.
	\end{itemize}
	Note that this notion also turns out to be useful for explicitly explaining how the semantics in the state-encoding CS calculus is related to request-response interactions. 

	Let us begin with the semantics of {\stateenccs} extended with session annotations where one session exactly corresponds to one request-response interaction on the web. 
\begin{center}
\begin{tabular}{l l}
 (Req)
 & $\llet{x}{\req{v_F}{\overline{w}}}{m} \ \overset{\sessionNothing}{|}\  \emp
 \ \ \ \Rightarrow^{enc} \ \ \ 
 \LetK{x}{m} \ \overset{\sessionSomething}{|} \ \llet{r}{v_F(\overline{w})}{r}$
 \\
 & 
 		\ \ \ where \  $v_F=\clo{F}{\bar{v}}$, \
 					 $\phi_\server(F)=\bar{z}\lamL{\server}{\bar{x}}{m_0}$, \ 
 					 	and \ fresh $session$
 \\
 (Call)
 & $\LetK{x}{m} \ \overset{\sessionSomething}{|} \ \call{v_F}{\overline{w}}
 \ \ \ \Rightarrow^{enc} \ \ \ 
 \llet{x}{v_F(\overline{w})}{m} \ \overset{\sessionNothing}{|} \ \emp$
 \\
 &
 	\ \ \ where \ $v_F=\clo{F}{\bar{v}}$, \  $\phi_\client(F)=\bar{z}\lamL{\client}{\bar{x}}{m_0}$
 \\
 (Reply)
 & $\LetK{x}{m} \ \overset{\sessionSomething}{|}v
 \ \ \ \Rightarrow^{enc} \ \ \ 
 \llet{x}{v}{m} \ \overset{\sessionNothing}{|} \ \emp$ 
\end{tabular}
\end{center}
(Req) corresponds to the creation of a new session. Both (Call) and (Reply) correspond to the closing of the session. For the other local rules, the status of the session remains unchanged:
\begin{center}
\begin{tabular}{l l l l}
Client & (AppC), (ValC), (LetC) & : & $m \ \overset{\ \ \ \sessionNothing \ \ \ }{|} \ \emp  \ \ \ \Rightarrow^{enc} \ \ \ m' \  \overset{\ \ \ \sessionNothing \ \ \ }{|} \ \emp$
\\
Server & (AppS), (ValS), (LetS) & : & $\pi \  \overset{\ \ \  \sessionSomething \ \ \ }{|} \ m \ \ \ \Rightarrow^{enc} \ \ \ \pi \ \overset{\ \ \ \sessionSomething \ \ \ }{|} \ m'$
\end{tabular}
\end{center}
where $m$, $m'$, and  $\pi$ are those from the semantic rules in {\stateenccs}.

	Now let us discuss the semantics of {\statefulcs} extended with session annotations where one session can not only correspond to one request-response interaction on the web but can also be expanded to correspond to more than one interaction: 
\begin{center}
\begin{tabular}{l l }
(Req1)
 & $\llet{x}{\req{v_F}{\overline{w}}}{m} \ \overset{\sessionNothing}{|} \ \emp
 \ \ \ \Rightarrow^{state} \ \ \ 
 \LetK{x}{m} \ \overset{\sessionSomething}{|} \ \emp; \ \llet{r}{v_F(\overline{w})}{r}$
 \\
 & \ \ \ where \ $v_F=\clo{F}{\bar{v}}$, \  $\phi_\server(F)=\bar{z}\lamL{\server}{\bar{x}}{m_0}$, \ and \ fresh $session$
 \\
(Req2)
 & $\llet{x}{\req{v_F}{\overline{w}}}{m} \ \overset{\sessionSomething}{|} \ \delta
 \ \ \ \Rightarrow^{state} \ \ \ 
 \LetK{x}{m} \ \overset{\sessionSomething}{|} \ \delta; \ \llet{r}{v_F(\overline{w})}{r}$
\\
 & \ \ \ where \ $v_F=\clo{F}{\bar{v}}$, \  $\phi_\server(F)=\bar{z}\lamL{\server}{\bar{x}}{m_0}$,
   \ and \ $\delta\not=\emp$
 \\
(Reply1) 
 & $\LetK{x}{m} \ \overset{\sessionSomething}{|} \ \emp; \ v
	\ \ \ \Rightarrow^{state} \ \ \ 
	\llet{x}{v}{m} \ \overset{\sessionNothing}{|} \ \emp$
 \\
(Reply2) 
 & $\LetK{x}{m} \ \overset{\sessionSomething}{|} \ \delta; \ v
 \ \ \ \Rightarrow^{state} \ \ \ 
 \llet{x}{v}{m} \ \overset{\sessionSomething}{|} \ \delta$
\ \ \ if $\delta\not=\emp$
 \\
(Call) 
 & $\LetK{y}{m_2}\ \overset{\!\!\sessionSomething\!\!}{|} \ \delta; \ \llet{x}{\call{v_F}{\overline{w}}}{m_1} 
   \Rightarrow^{state} 
 \llet{y}{v_F(\overline{w})}{m_2} \ \overset{\!\!\sessionSomething\!\!}{|} \ \LetK{x}{m_1}\cdot\delta$
 \\
 & \ \ \ where $v_F=\clo{F}{\bar{v}}$ \ and \ $\phi_\client(F)=\bar{z}\lamL{\client}{\bar{x}}{m_0}$
 \\
(Ret) 
 & $\llet{y}{\ret{v}}{m_2} \ \overset{\sessionSomething}{|} \ \LetK{x}{m_1}\cdot\delta 
	\ \ \ \Rightarrow^{state} \ \ \ 
 \LetK{y}{m_2} \ \overset{\sessionSomething}{|} \ \delta; \ \llet{x}{v}{m_1}$
\end{tabular}
\end{center}

	The original semantic rule (Req) in {\statefulcs} is divided into two rules. (Req1) with the empty server stack corresponds to the creation of a new session, and (Req2) with the non-empty stack $\delta$ corresponds to a continuation of the previously created session. The original semantic rule (Reply) in {\statefulcs} is also divided into two rules. (Reply1) with the empty server stack is matched with (Req1) to close the created session, and (Reply2) with the non-empty stack $\delta$ is matched with (Req2) to maintain the previously created session. Both (Call) and (Ret) are extended to continue to maintain the status of the session. For the other local rules, the status of session remains unchanged as 

\begin{center}
\begin{tabular}{l l l l}
Client & (AppC), (ValC), (LetC) & : & $m \overset{\ \ \ \sessionOption \ \ \ }{|} \delta \ \ \
 \Rightarrow^{enc} \ \ \ m' \ \overset{\ \ \ \sessionOption \ \ \ }{|} \ \delta$ 
\\
Server & (AppS), (ValS), (LetS) & : & $\pi \overset{\ \ \ \sessionSomething \ \ \ }{|} \delta; m \ \ \
 \Rightarrow^{enc} \ \ \ \pi \overset{\ \ \ \sessionSomething \ \ \ }{|} \ \delta; m'$
\end{tabular}
\end{center} 
where $m$, $m'$, $\pi$, and $\delta$ are those from the semantic rules in {\statefulcs}. For the client-side local rules, it is confirmed that either $optSession=nothing$ and $\delta=\emp$ or $optSession=session$ and $\delta\not=\emp$ holds. 

	The session-annotated semantics of {\statefulcs} clearly explains how the stateful CS calculus using request-response interactions is implemented on the web-based model. Also, the idea of applying session annotation to the semantics enables us compare {\stateenccs} with {\statefulcs} in terms of the web-based implementation using request-response interaction.

\begin{figure}[t]
\centering
\includegraphics[width=0.5\textwidth]{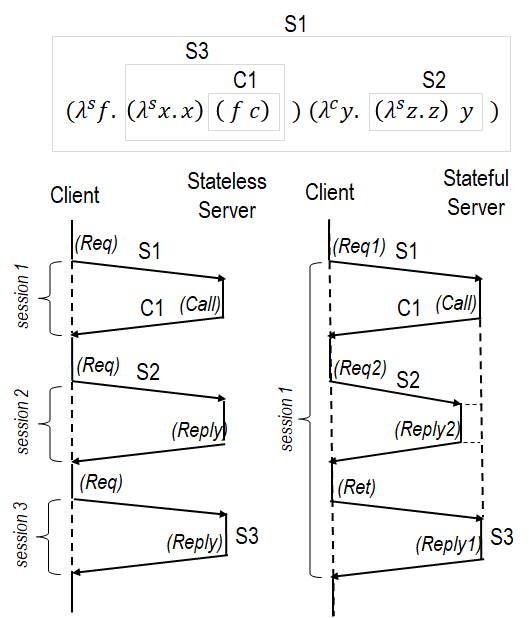}
\caption{Comparison of session management between {\stateenccs} and {\statefulcs}}
\label{fig:statelessvsstatefulrpcexample}
\end{figure}

Figure \ref{fig:statelessvsstatefulrpcexample} shows a comparison of session management between {\stateencrpc} and {\statefulrpc} when the {\rpc} term in the example is evaluated. First, the figure shows that client-server communication in {\statefulrpc} is the same as that in {\stateencrpc} by using the same box names S1, C1, and S2 on the related flows. Second, with the stateful server, the active thread of control described by the solid line with (S2) is above the pending thread of control described by the dashed line with (S3), meaning that S2 is being evaluated with the server stack holding the server context of S3. 

\subsubsection{A Mixed Strategy}

	In the previous comparison between {\stateenccs} and {\statefulcs}, we clearly
understood how they manage sessions differently. 
	In {\stateenccs}, the server does not need to retain anything once it sends the client a response. This is advantageous because the server resource consumption is reduced. 
	In {\statefulcs}, the server can naturally support  storing a stateful context that the server must have before and after a client function call. This is necessary when it is not easy or not possible to encode the stateful context, for example, in database operations as seen in Section \ref{sec:statefulcalculi}, and to send the encoded context to the client. 
	The question then arises if we can use both strategies: a mixed strategy basically employs the state-encoding strategy to reduce the server resource consumption but switches to the stateful strategy when necessary. We here argue that our theory can be extended to support this mixed strategy. 
	
	For the mixed strategy, the problem is how to distinguish one phase of {\rpc} terms in which the compilation rules and semantic rules of {\stateencrpc} and {\stateenccs} are applied, from the other phase of the terms in which those of {\statefulrpc} and {\statefulcs} are applied. A simple idea is to use the notion of monadic encapsulation of state \cite{Launchbury1994,Timany:2017}, where stateful computations are encapsulated using monads and the purity of the functional language embracing them is preserved. 

	To explain how monadic encapsulation of state can help in the design of our mixed strategy, we first introduce the standard monadic operations and stateful operations to the term syntax of {\rpc}, and a new (state) type is introduced to the type syntax:
	\[
	\begin{array}{l c l}
		M &::=& \cdots \ | \ runST \ M \ | \ thenST \ M \ (\lambda x.M) \ | \ returnST \ M \\
		  &|& newVar \ M \ | \ readVar \ M \ | \  writeVar \ M \ M \\
		\tau &::=& \cdots \ | \ ST \ state \ \tau \ | \ \forall state. \tau
	\end{array}
	\]
	A key construct $runST \ M$ encapsulates a stateful computation $M$ where states may be cascaded through $newVar$, $readVar$, and $writeVar$ by the monadic binder $thenST$, which allows a value to escape from the monad by the monadic stopper $returnST$ at the end. 
	Note that $state$ is a (state) type variable, which is essential to the monadic encapsulation of state in a pure functional language. A new state type $ST \ state \ \tau$ can be read intuitively as follows: a stateful computation of this type can allocate, read, and write in a memory region indexed by $state$ and then produce a value of type $\tau$. The type of $runST$ in the literature \cite{Launchbury1994,Timany:2017} is 
	\[
	(\forall state. \ ST \ state \ \tau) \rightarrow \tau.
	\]
Having a universally quantified state type variable in the type guarantees that every memory region allocated inside the stateful computation is not accessible outside the $runST$ anymore by limiting the scope of the state type variable, which never occurs in $\tau$. This is a key property of the monadic encapsulation of state.
	
	For example, in an evaluation of the following extended {\rpc} term: 
	\[
	\begin{array}{l}
	clientFun2 \ ( \\
	\ \ \ runST \ ( \\
	\ \ \ \ \ \ (thenST \ (newVar \ 0) \ (\lambda l. \\
	\ \ \ \ \ \ (thenST \ (writeVar \ l (clientFun1 ())) \ (\lambda (). \\
	\ \ \ \ \ \ (thenST \ (readVar \ l) \ (\lambda x. \\
	\ \ \ \ \ \ (returnST x))))))) \ \ \ ) \ )
	\end{array}
	\]
	a new piece of memory is allocated	at an address $l$ and is initialised with 0 by $newVar \ 0$, 
	 the server calls a client function to obtain a value by $clientFun1()$, 
	the memory is updated with the value by $writeVar \ l \ (\cdots)$, 
	the written value is read again by $readVar \ l$, and 
	it escapes the $runST$ monad by $returnST \ x$. Finally, the other client function $clientFun2$ is called from the server with the escaped value as an argument. 
	
	It is important to note that $clientFun1$ is called in the middle of the stateful computations, whereas $clientFun2$ is called after all the stateful computations. Therefore, the former should be implemented in the stateful server strategy, but the latter can be implemented in the state-encoding server strategy.  
	To capture these different phases, we introduce two forms of typing judgements as follows:
	\[
	\typing{\Gamma}{a}{M}{\tau} \ \ \ \ \ \ \ \ \ \ \typingBlack{\Gamma}{a}{M}{\tau}
	\]
where the typing judgements with the black triangle denote that a term under evaluation is in the middle of stateful computations, and the typing judgements with the white triangle denote the absence of stateful computations. We also change the function type to include the colour as $\tau \funL{a, \ colour} \tau'$, where it is white or black. Functions of the black function type are invoked in the middle of the stateful computations, whereas functions of the white function type are never invoked there. 

	The typing rules for {\rpc} are then duplicated, and therefore we make one set have the typing rules with only the white triangle and make the other set have them with only the black triangle. In (T-App-B) of the black set, the colour of the function type in the premise is made the same as the colour of the triangle. In (T-Lam-B), the colour of the function type in the conclusion is the same as the colour of the triangle.
	These coloured typing rules will {\it get a function infected} whenever the function is invoked within a stateful expression. 
\[
	\begin{prooftree}
		\hypo{ \typingBlack{\tyenv}{a}{L}{\tau\funL{a,black}\tau'} }
		\hypo{ \typingBlack{\tyenv}{a}{M}{\tau}  }
		\infer[left label=(T-App-B)]2{ \typingBlack{\tyenv}{a}{L \ M}{\tau'} }
	\end{prooftree}
\ \ \ \ \ 
	\begin{prooftree}
		\hypo{ \typingBlack{\tyenvExt{x}{\tau}}{b}{M}{\tau'} }
		\infer[left label=(T-Lam-B)]1{ \typingBlack{\tyenv}{a}{\lamL{b}{x}{M}}{\tau\funL{b,black}\tau'} } 
	\end{prooftree}
\]
	
	The two sets of the typing rules are combined through two new typing rules:
\[
	\begin{prooftree}
		\hypo{  fresh \ state, \ \ state \not\in FTV(\tyenv) \cup FTV(\tau) }
		\hypo{  \typingBlack{\tyenv}{\server}{M}{ST \ state \ \tau} }
		\infer[left label=(Gen\&Run)]2{ \typingBlack{\tyenv}{\server}{runST \ M}{\tau}   }
	\end{prooftree}
\]
\[
	\begin{prooftree}
		\hypo{ \ FTV(\tyenv) \cup FTV(\tau) = \emptyset }
		\hypo{ \typingBlack{\tyenv}{\server}{runST \ M}{\tau}  }
		\infer[left label=(Purify)]2{  \typing{\tyenv}{\server}{runST \ M}{\tau} }
	\end{prooftree}
\]
(Gen\&Run) is a typing rule combining (Gen) and (Run) in the literature, which generalises a state type variable and is a typing rule for $runST$, respectively. Those typing rules for the monadic constructs and state manipulation constructs are from the literature, and they are included in the black set. (Purify) is a typing
rule that encapsulates a stateful world from pure functional expressions whenever no free state type variables escape, i.e. $FTV(\tyenv) \cup FTV(\tau) = \emptyset$.

	With these typing rules, $clientFun1$ in the example must be typed by (T-App-B); it {\it gets infected} because it is invoked within a runST expression, which is typed by (Gen\&Run). However, $clientFun2$ can be typed by (T-App-W), which is a white-coloured version of (T-App-B). The scope of states of the runST expression is thus limited inside it by (Gen\&Run). Then it can be treated as a pure expression by (Purify) because there are no other runST expressions enclosing it. 
	
	After typing an extended {\rpc} term with the coloured typing rules, the phases using the white typing rules are compiled and evaluated under the state-encoding server strategy, and the phases using the black typing rules are compiled and evaluated under the stateful server strategy. Monadic encapsulation of state has been proved to separate the stateful world from the pure expression in the literature \cite{Launchbury1994,Timany:2017}. Therefore, we can utilise the idea to separate where we can use the state-encoding server strategy from where we should use the stateful server strategy. What remains to be seen is how to compile (Purify). This typing rule is compiled as a direct style invocation of a CPS-style expression compiled from the typing in the premise by applying the initial continuation to the expression.
	This can be viewed as a border between the two strategies. 
	
	This completes the explanation of how we designed a hybrid strategy, which is another benefit of our theory of RPC calculi. 

\section{Related Work}
\label{sec:relatedwork}

	Links \cite{Cooper:2006:LWP:1777707.1777724} is a practical multi-tier web programming language that employs the RPC calculus as the foundation for client-server communication. It is a typed functional programming language incorporating Kleisli-based database query optimisation, continuations for web interactions, concurrency with message passing, and XML programming. It would be interesting to extend the implementation of Links using the proposed theory of RPC calculi. 
	
	Lambda5 \cite{MurphyVII:2004:SML:1018438.1021865,Murphy:2008:MTM:1467784} is a modally-typed lambda calculus in which modal type systems can control local resources safely in distributed systems. It
	is extended to a full-fledged web programming language. From the programmer's perspective, there are two things to compare. First, while Lambda5 is similar to the RPC calculus in that it allows programmers to write symmetric communication, it distinguishes the syntax of remote function applications, say, `$\texttt{get}\langle\server\rangle(\texttt{here} \ M \ N)$', from that of local ones, say, `$M \ N$'. In this respect, Lambda5 is like {\stateencrpc} and {\statefulrpc} because {\rpc} offers
	a uniform syntax for both kinds of applications. 
	
	Second, the modal type system for Lambda5 introduces \texttt{at} modality, `$\tau \ \texttt{at} \ a$' in order to be able to refer to a specific location. The use of \texttt{at}-types for some term $M$ and typing environment $\tyenv$ can be viewed as typing judgements $\typing{\tyenv}{a}{M}{\tau}$. One interesting difference is that our type system introduces locative types $\tau\funL{a}\tau'$ with specific location $a$. The type system for Lambda5 uses box types $\Box\tau$ and diamond types $\Diamond\tau$, but locations in the modal types are too implicit to be usable; box types are associated with all arbitrary locations, and diamond types are associated with some unknown location. On the basis of this observation, we believe that our type system is a modal type system using locative types that internalise \texttt{at} modality. A formal comparison on how our type system is related to the modal type system is left as a future work. 
	Another interesting question is to investigate if the state-encoding and stateful calculi  could have any modal type systems beyond the type system for the RPC calculus.
	
	From the implementation perspective, Lambda5 designed for peer-to-peer distributed systems is based on symmetric communication, whereas the RPC calculus designed for client-server systems is based on asymmetric communication. The semantic rules for client and server in Lambda5 are therefore the same. Although Lambda5 could be exploited for a stateless server strategy, the same semantic rules have to be revised in a nontrivial way to achieve a stateful client and stateless server, where each is defined by different semantic rules. For a stateful server strategy, the Lambda5 semantic rules should also be revised to make them usable with the client-server model. The idea used in Lambda5 of continuations spanning multiple worlds such as client and server is similar. However, in {\statefulrpc}, the semantic rules (Req) and (Call) for client-to-server RPCs and vice versa are asymmetric; the former is defined to initiate a new server control context, and the latter is defined to return to a client context waiting for this server response. This fits the implementation based on trampolined-style well. Thus, the semantics of Lambda5 has this subtle gap between its description and its web-based implementation.
	Lambda5 is implemented on a web-based client-server model in an ad hoc way, for example to support channels on top of HTTP.  
	
	A multi-tier calculus by Neubauer and Thiemann \cite{Neubauer:2005:SPM:1040305.1040324} automatically constructs communications for concurrently running processes. The calculus employs session types \cite{Gay:1999:TSC:645393.756448} to enforce the integrity of communications automatically. They proposed a series of transformations as compilation to convert a source program into separate programs at different locations determined by the use of primitives that run only at specific locations. The calculus can be called a {\it stateful peer strategy} because each process manages its own state, but there is no concept of a central server as in Lambda5. 

	There are several programming languages that have adopted the feature of multi-tier web programming: 
Hop \cite{Serrano2006, Serrano:2012:MPH:2240236.2240253} extending Scheme; Hop.js extending JavaScript \cite{Serrano:2016:GH:3022670.2951916};	Eliom, a multi-tier ML programming language, featuring module systems extended with location annotations \cite{Radanne2017} in the project Ocsigen \cite{Balat2006}; and Ur/Web \cite{Chlipala2015} with a dependently typed system; and a multi-tier functional reactive programming framework for Scala \cite{Reynders2014}. They are all practical web programming languages featuring concurrency, reactive UI, database, and XML. 

	Hop, Eliom, and Ur/Web provide programmers only asymmetric communication where basically only the client can invoke server functions; the server resorts to a network library to call client functions back. All of them use a special syntax for RPC distinct from that for local function calls. 
	As far as we know, they are all based on a stateful server approach as described for HOP in \cite{Serrano2010,Boudol2012} and for Eliom in \cite{Radanne2017}, though Ur/Web never presented any formal semantics on their RPC implementation. 
	These languages do not need to implement the mapping of symmetric communication onto asymmetric communication as is done in our theory. It is therefore less meaningful to compare our theory with them in terms of stateful server approaches.  
	In \cite{Serrano2010}, a denotational continuation-based semantics for a core subset of HOP is present for the purpose of elaboration of how to generate client-side code by the server-side code. In \cite{Boudol2012}, more versatile semantics is proposed in order to reason about the safety of the same-origin policy in the web application security model, whereas we use our semantics for proving the correctness of compilation for the two RPC implementation strategies. 
 
	The PLT Scheme community has studied the construction of interactive server-side web applications.
	In \cite{Graunke:2001:ARP:872023.872573,Matthews2004AutomaticallyRP}, the authors have pointed out the mismatch problem between the structure of {\it sequential} programs and the structure of the corresponding web interactions. They suggest an automatic structuring method of sequential programs so that a restructured web program and a consumer can construct a pair of coroutines where each interaction point can be resumed arbitrarily often. The method stores a continuation object at the server, referring to it by ID, which can be viewed as a stateful server strategy. In the implementation of our stateful calculi, a client and a server play exactly the same role as coroutines, as explained in Section \ref{sec:compforstatefulcs}, though we use session numbers sent from the client on requests to resume the suspension. 
In \cite{McCarthy:ICFP2009,McCarthy:OOPSLA2010,Krishnamurthi:2007:IUP:1325174.1325176}, the authors have addressed the practical problem of how scalable, continuation-based web programs can achieve modular compilation and higher-order interaction with third-party libraries by stack inspection and manipulation techniques known as continuation marks and delimited continuations. Although the techniques used are different, their solution can be viewed as a mixed strategy from the standpoint of our theory. 


	Using continuations as a language-level technique to structure the client-server interactions of web programs is well-known, as has been discussed by \cite{Queinnec2004}. 
	Although it is not about multi-tier web programming, it is worth noting that this work discusses the advantages and disadvantages of storing continuations on the client side and  server side. 

	Wysteria \cite{Rastogi:2014:WPL:2650286.2650775} is another related work. This system is in the different domain of secure multi-party computation but shares some similar technical machinery and development. It indexes the types of data and computations by participating principals, which can be regarded as locations. Further, Wysteria also proposes abstractions for private states among participating principals and among subsets of these principals, acting as a secure computation. Similar to our research, it also proposes a translation whose targets consists of `sliced' code for each participant.

\section{Conclusion}
\label{sec:conclusion}

	We have proposed a theory of RPC calculi that deals with the implementation of arbitrarily deep nested client-server interactions using the feature of symmetric communication on the web-based flat request-response interactions in trampolined style under the state-encoding
	server strategy and the stateful server strategy. In this theory, a new typed RPC calculus {\rpc} is proposed where typing with locative types identifies remote function applications at the type level. 
	Our new theory of RPC calculi has advantages over the existing research work. With the help of type-level identification of remote functions, the structure of compilation of {\rpc} into {\stateencrpc} and {\statefulrpc} is simplified compared to the structure of the original compilation without using type information. This simplicity allows our theory to explore three design choices: a state-encoding server strategy, a stateful server strategy, and furthermore, a mixed strategy of these two strategies. 
	
	In future, we hope to design a full-fledged multi-tier web programming language based on our theory of RPC calculi. The design must consider many more features such as HTML constructs, reactive UI, databases, and concurrency because our theory addresses only client-server communication. 

\bibliographystyle{jfp}

\appendix

\section{A Proof of Theorem for the RPC Calculus} 


\begin{theorem}[Type Soundness for \rpc] If \ $\typing{\tyenv}{a}{M}{\tau}$ and $\evalRPC{M}{a}{V}$ then $\typing{\tyenv}{a}{V}{\tau}$.
\end{theorem}
\begin{proof}
We prove this theorem by induction on the height of the typing derivation on $M$. For \mbox{(T-Var)} as a base case, it is provable trivially. For inductive cases, we do case analysis on the last typing rule used. For (T-Lam), it is straightforward. For (T-App), (T-Req), and (T-Call), we use inductive arguments and the substitution lemma \ref{lem:substitution}. In order to prove the cases with (T-Req) and (T-Call), we need an additional fact \ref{fact:locationindependence} saying that the types of values are the same wherever the values are.
\end{proof}

\begin{fact} If $\typing{\tyenv}{a}{V}{\tau}$ then $\typing{\tyenv}{b}{V}{\tau}$.
\label{fact:locationindependence}
\end{fact}
\begin{proof}
This is provable due to the definition of (T-Var) and (T-Lam) which are typing rules for $V$, either a variable and a function. 
\end{proof}

\begin{lemma}[Substitution] If $\typing{\tyenv,x:\tau_1}{a}{M}{\tau_2}$ and $\typing{\tyenv}{b}{W}{\tau_1}$ then $\typing{\tyenv}{a}{M\{W/x\}}{\tau_2}$.
\label{lem:substitution}
\end{lemma}
\begin{proof}
The proof is straightforward by induction on the height of the typing derivation on $M$.
\end{proof}

\section{Proofs of Theorems for the State-Encoding Calculi} 


The main theorem states the correctness of compilation for {\stateencrpc} as follows.
\begin{theorem}[Correctness of Compilation for \stateencrpc] Assume a well-typed {\rpc} term $M$ under the locative type system.
\label{thm:correctcompstateencrpc}
\begin{itemize}
\item If $\evalRPCC{M}{V}$ then \runStateEncCSStar{\ccomp{M} \ | \ \emp}{\ccomp{V} \ | \ \emp}.
\item If $\evalRPCS{M}{V}$ then \runStateEncCSStar{\Pi \ | \ \scomp{M} \ K}{\Pi \ | \ \scomp{V} \ K}  for all $\Pi$ and $K$.
\end{itemize}
\end{theorem}
\begin{proof}
	The proof consists of two parts. The proof firstly builds a sequence of evaluation of the compiled term in {\stateencrpc} matched to each subtree of evaluation of a given term in {\rpc} by induction on the height of the evaluation of $M$. Then the proof puts the sequences together to construct a longer one matched to the whole evaluation. The composition lemma, Lemma \ref{lem:evalpreserveletstateencrpc}, supports this by showing that, given an  {\stateencrpc} evaluation sequence, it is possible to construct the same sequence but under arbitrary let bindings. The proof uses this lemma as glue in the client. But for the same purpose in the server, the lemma is not used because the proof uses continuations as glue.
 
	In addition, the proof must derive the use of substitutions in {\stateencrpc} from the use of those in {\rpc} to complete the construction of the {\stateencrpc} evaluation sequence. The substitution lemma, Lemma \ref{lem:substitutionstateencrpc}, supports this by showing that a compilation of a substituted term in {\rpc} is the same as a substitution in {\stateencrpc} of compiled terms.

	 In the RPC calculus, every base case involves only (Value) where the height is 1, and every inductive case must use (Beta) to make the height at least higher than 1. For the base cases with (Value), both of $M$ and $V$ are $\lamL{b}{x}{M_0}$. The theorem can be easily verified since $\ccomp{M}=\ccomp{V}$ is true. 
	
	In the inductive cases, $M$ is $\appL{L}{}{M_{arg}}$. The inductive cases are also analysed into four by the similar reason since (Beta) involves two locations $a$ and $b$: the application is being evaluated at the location $a$, and $L$ evaluates to a function of the location $b$. 
	
	i)
	When $a$ is $\server$: Firstly, by the induction hypotheses with three subtrees in the evaluation of the {\rpc} term $M$, three subsequences in {\stateencrpc} can be constructed with universally quantified variables $K_1,K_2,K_3$. Secondly, to combine these subsequences in order to make what the theorem demands, the proof instantiates the three variables with a continuation to $\scomp{L}$ (see Figure \ref{fig:compforstateencrpc}), a continuation to $\scomp{M_{arg}}$ (see Figure \ref{fig:compforstateencrpc}), and $K$ in the compilation rules (AppSS) or (AppSC) at Figure \ref{fig:compforstateencrpc}, respectively. Note that the proof for these inductive cases ($a$ is $\server$) is thus proved without using Lemma  \ref{lem:evalpreserveletstateencrpc}, which is necessary for the following inductive cases ($a$ is $\client$). Finally, the proof uses Lemma \ref{lem:substitutionstateencrpc} to complete building the evaluation sequence in {\stateencrpc} as the theorem demands.  
	
	ii)
	When $a$ is $\client$: This proof firstly uses the induction hypotheses with the three subtrees in the evaluation of the {\rpc} term $M$ to build {\stateencrpc} evaluation subsequences. Secondly, the proof has to combine these subsequences in order to form a longer one that the theorem demands. Each of $\ccomp{L}$, $\ccomp{M_{arg}}$, and a local or remote application is surrounded by a let term of the form $\llet{var}{\cdots}{term}$ in the compilation rules (AppCC) and (AppCS) at Figure \ref{fig:compforstateencrpc}. So, the proof has to resort to Lemma \ref{lem:evalpreserveletstateencrpc} to lift an evaluation subsequence for $\ccomp{L}$ to one for $\llet{var}{\ccomp{L}}{term}$, and to lift another evaluation subsequence for $\ccomp{M_{arg}}$ to one for $\llet{var'}{\ccomp{M_{arg}}}{term'}$. Finally, the proof takes the third evaluation subsequence for the local or remote application to finish by using Lemma \ref{lem:substitutionstateencrpc}.
\end{proof}

	To prove Theorem \ref{thm:correctcompstateencrpc}, we use Lemma \ref{lem:evalpreserveletstateencrpc} and Lemma \ref{lem:substitutionstateencrpc}.
	The former lemma states that the evaluation of terms in {\stateencrpc} is preserved under the let binding. 

\begin{lemma}[Composition] In the state-encoding RPC calculus {\stateencrpc},
\label{lem:evalpreserveletstateencrpc}
\begin{itemize}
\item If \runStateEncCSStar{M \ | \ \emp}{V \ | \ \emp} \\
then \runStateEncCSStar{\llet{x}{M}{M_0} \ | \ \emp}{\llet{x}{V}{M_0} \ | \ \emp}
\end{itemize}
\end{lemma}
\begin{proof}
	This is proved by induction on the length of the evaluation in the condition. In the base case, the length is zero, and so $M$ is identical to $V$. Therefore, $\llet{x}{M}{M_0}$ is identical to $\llet{x}{V}{M_0}$, which proves the base case. 
	
	In inductive cases where the length is one or longer, the proof does a case analysis on kinds of the first rule used in the evaluation of \runStateEncCSStar{M \ | \ \emp}{V \ | \ \emp} and \runStateEncCSStar{\Pi \ | \ M}{\Pi \ | \ V}, to make the evaluation progress by applying a rule for the kind. Then the proof will have an evaluation subsequence shorter than one in the condition, with which we can apply induction to finish the proof. 
	
	 Hence, there are 4 cases for consideration.
	
	i) (AppC) and (ValC): The lemma is provable simply by induction as explained above. 

	ii) (Req): Let $M$ be $\llet{y}{\req{V_f}{\bar{W}}}{M_y}$. Then there exist steps such that
	
\begin{center}
\begin{tabular}{l l l}
                 & $\llet{y}{\req{V_f}{\bar{W}}}{M_y} \ | \  \emp$ \\
$\RightarrowEnc$ & $\LetK{y}{M_y} \ | \ V_f(\bar{W})$ \\
$\RightarrowEnc$ & $Conf_1 \RightarrowEnc ... \RightarrowEnc Conf_k \RightarrowEncStar V \ | \  \emp$
\end{tabular}
\end{center}
where either (Reply) or (Call) must be used in some step, $Conf_{k-1}\RightarrowEnc Conf_k$, and neither (Reply) nor (Call) is used before the step. This is because, in order to move from the server configuration $(\LetK{y}{M_y} \ | \ V_f(\bar{W}))$ to a client configuration $(V \ | \ \emp)$, there is no other way than using (Reply) or (Call) by the definition of the semantic rules for {\stateencrpc}. 

	After the use of either (Reply) or (Call), it is easy to construct, 
	\[
	Conf_k \RightarrowEncPlus Conf_{k+n} \RightarrowEncStar V \ | \  \emp
	\]
	where all except $Conf_{k+n}$ are for the server and $Conf_{k+n} \RightarrowEncStar V \ | \  \emp$ is a condition that we can apply induction with. This finishes the proof for this case.
	
		iii) (LetC): In this case, $M$ is $\llet{x_1}{(\llet{x_2}{M_1}{M_2})}{M_3}$. This is  proved by induction after using the intermediate value lemma, Lemma \ref{lem:letbindingfirstonestateencrpc}, and the context-independent intermediate value lemma, Lemma \ref{lem:letbindingfirsttwostateencrpc}. By applying (LetC) two times, we have:
		\[
		\llet{x}{M}{M_0} \ | \ \emp \ \ \ \Rightarrow^2 \ \ \ 
		\llet{x_2}{M_1}{(\llet{x_1}{M_2}{(\llet{x}{M_3}{M_0})})} \ | \ \emp
		\]
		where induction cannot be applied immediately.
		After the evaluation of both $M_1$ and $M_2$, we will reach a configuration as:
		$
		\llet{x}{M_3\{V_1/x_1\}}{M_0} \ | \ \emp
		$
		where $V_1$ is a value of $(\llet{x_2}{M_1}{M_2})$. Then we can apply induction with 
		$ M_3\{V_1/x_1\} \ | \ \emp \Rightarrow^* V \ | \ \emp
		$.
		We thus need to make the evaluation progress over $M_1$ and $M_2$. 
		To do this, we apply the two additional lemmas to lead the evaluation to the configuration above. This finishes the proof in the case. 

\end{proof}

	Lemma \ref{lem:substitutionstateencrpc} shows the compilation of a substituted term yields the same term as the compilation of a term substituted with compiled values. 

\begin{lemma}[Substitution] The compilation $\ccomp{-}$ and $\scomp{-}$ of {\rpc} into {\stateencrpc} preserve substitution operations in {\rpc} and {\stateencrpc}.
Let $\vcomp{\lamL{a}{x}{M}}=\ccomp{\lamL{a}{x}{M}}$.
\label{lem:substitutionstateencrpc}
\begin{itemize}
\item $\ccomp{N\{W/x\}} = \ccomp{N}\{\vcomp{W}/x\}$.
\item $\scomp{N\{W/x\}} \ K = (\scomp{N} \ K)\{\vcomp{W}/x\}$.
\end{itemize}
\end{lemma}
\begin{proof}
This lemma is true because of the definition of the compilation rules. 
\end{proof}

To prove Lemma \ref{lem:evalpreserveletstateencrpc}, we introduce Lemma \ref{lem:letbindingfirstonestateencrpc} and Lemma \ref{lem:letbindingfirsttwostateencrpc}, as follows.
\begin{lemma}[Intermediate Value]
\label{lem:letbindingfirstonestateencrpc}
\begin{itemize}
\item If \runStateEncCSStar{\llet{x}{M}{M_0} \ | \ \emp}{V \ | \ \emp} \\
then $\llet{x}{M}{M_0} \ | \ \emp 
		\Rightarrow^{enc*} \llet{x}{V_x}{M_0} \ | \ \emp 
		\Rightarrow^{enc*} V \ | \ \emp$.
\end{itemize}
\end{lemma}
\begin{proof}
This is proved by induction on the length of the evaluation in the conditions.
\end{proof}

\begin{lemma}[Context-Independent Intermediate Value]
\label{lem:letbindingfirsttwostateencrpc} 
\begin{itemize}
\item If \runStateEncCSStar{\llet{x}{M}{M_1} \ | \ \emp}{\llet{x}{V}{M_1} \ | \ \emp} \\
then \runStateEncCSStar{\llet{x}{M}{M_2} \ | \ \emp}{\llet{x}{V}{M_2} \ | \ \emp}.
\end{itemize}
\end{lemma}
\begin{proof}
This is proved by induction on the length of the evaluation in the conditions. In (ValC), (AppC), and (Req), the proof is straightforward. In (LetC), when  $M$ is $(\llet{y}{M_y}{M_x})$, we have $(\llet{x}{M}{M_1})\ |\ \emp \Rightarrow^{enc}\llet{y}{M_y}{(\llet{x}{M_x}{M_1})}\ |\ \emp$. The proof uses an inner induction on the number of the let constructs enclosing $\llet{x}{M_x}{M_1}$ such as ``$\llet{y}{M_y}{-}$''. The use of the inner induction will eventually lead to a form of $\llet{x}{M_x\{V_y/y\}}{M_1} \ | \ \emp$  where $V_y$ is a value of $M_y$. Then we can apply the outer induction to it to finish the proof.
\end{proof}

The theorem of the correctness of compilation for {\stateenccs} immediately follows the theorem for {\stateencrpc}. 
\begin{theorem}[Correctness of Compilation for \stateenccs] Assume a well-typed term $M$ under the locative type system. Given $\phi_\client$ and $\phi_\server$, 
\label{thm:correctcompstateencs}
\begin{itemize}
\item If $\evalRPCC{M}{V}$ then 
\runStateEncCSStar{\cconv{\ccomp{M}} \ | \ \emp}{\cconv{\ccomp{V}} \ | \ \emp}.
\end{itemize}
\end{theorem}
\begin{proof}
By Theorem \ref{thm:correctcompstateencrpc}, it holds that if $\evalRPCC{M}{V}$ then \ \runStateEncRPCStar{\ccomp{M} \ | \ \emp}{\ccomp{V} \ | \ \emp}. From such a series of evaluation steps in {\stateencrpc}, it is easy to construct evaluation steps in {\stateenccs} as \runStateEncCSStar{\cconv{\ccomp{M}} \ | \ \emp}{\cconv{\ccomp{V}} \ | \ \emp}. This is because of two reasons, as follows.

	First, the compilation rules $\cconv{-}$ preserves the structure of terms, for example, mapping a variable in {\stateencrpc} into a variable in {\stateenccs}, an application in {\stateencrpc} into an application in {\stateenccs}, and so on. The only exception is to map a function in {\stateencrpc} into a closure in {\stateenccs}, but a closure has a role as a function in {\stateenccs}. 
	Second, the semantic rules in {\stateencrpc} are isomorphic to the semantic rules in {\stateenccs}. 
	
	To construct evaluation steps starting from $\cconv{\ccomp{M}} \ | \ \emp$ in {\stateenccs}, we have only to choose the isomorphic semantic rule corresponding to the semantic rule used in each of the evaluation steps in {\stateencrpc}.
\end{proof}

\section{Proofs of Theorems for the Stateful Calculi} 


The structure of proofs of theorems for the stateful calculi is very similar to that of those for the state-encoding calculi. 

	The main theorem states the correctness of compilation for {\statefulrpc} as follows.
\begin{theorem}[Correctness of Compilation for \statefulrpc] Assume a well-typed term {\rpc} $M$ under the locative type system.
\label{thm:correctcompstatefulrpc}
\begin{itemize}
\item If $\evalRPC{M}{c}{V}$ then \runStatefulCSStar{\ccomp{M} \ | \ \Delta}{\ccomp{V} \ | \ \Delta} for all $\Delta$, which is call-return balanced.
\item If $\evalRPC{M}{s}{V}$ then 
\runStatefulCSStar{\Pi \ | \ \Delta; \scomp{M}}{\Pi \ | \ \Delta; \scomp{V}} for all $\Pi$ and $\Delta$, which is call-return balanced.
\end{itemize}
\end{theorem}
\begin{proof}
	This proof has the same structure as for Theorem \ref{thm:correctcompstateencrpc}. It firstly builds a subsequence of evaluation of the compiled term in {\statefulrpc} to each subtree of evaluation of a given term in {\rpc}, and then it puts the subsequences together into a longer one for the whole evaluation. The composition lemma, Lemma \ref{lem:evalpreserveletstatefulrpc}, supports the second part. The substitution lemma, Lemma \ref{lem:substitutionstatefulrpc} is required in the second part to show that the a {\statefulrpc} substitution after a {\rpc}-to-{\statefulrpc} compilation produces the same term as one by a {\rpc}-to-{\statefulrpc} compilation after a {\rpc} substitution. 

	Now let us prove by induction on the height of the evaluation of $M$. 	
	The base cases with height 1 use (Value) in the RPC calculus. For base cases with (Value), both $M$ and $V$ are the same as $\lamL{b}{x}{M_0}$, and so the theorem is trivially true by zero step evaluation.
	
	The inductive cases with higher than height 1 must use (Beta) in the condition of this theorem. Every case is provable by induction and the use of Lemma \ref{lem:evalpreserveletstatefulrpc} and \ref{lem:substitutionstatefulrpc} in the same way as for Theorem \ref{thm:correctcompstateencrpc}. 
	For the case with $\evalRPC{\appL{L}{\client}{M_{arg}}}{\client}{V}$, each evaluation subsequence in {\statefulrpc} obtained from each subtree of evaluation in {\rpc} is balanced by induction. The three subsequences are combined by having a single configuration using (ValC) between the first and the second, and by having another configuration using (AppC) between the second and the third. (ValC) is used in the end. By the definition of call-return balance, the resulting evaluation sequence in {\statefulrpc} is call-return balanced. The case with $\evalRPC{\appL{L}{\server}{M_{arg}}}{\server}{V}$ is verified exactly in the same way but with (ValS), (AppS), and (ValS). 
	
	For the case with  $\evalRPC{\appL{L}{\server}{M_{arg}}}{\client}{V}$, the three call-return balanced evaluation subsequences are combined with intermediate configurations using (ValC) between the first and the second, using (Req) and (Apps) between the second and the third, and using (ValS), (Reply), and (ValC) in the end.
	
	The case with $\evalRPC{\appL{L}{\client}{M_{arg}}}{\server}{V}$ is most interesting because a new pair of call and ret will be added to the resulting evaluation sequence explicitly, not just by induction. The first subsequence is followed by configurations using (ValS), which is followed by the second subsequence, which is followed by other configurations using (Call), (AppC), (LetC), and (AppC), which is followed by the third subsequence, which is lastly followed by a configuration using (Ret). The whole evaluation sequence is call-return balanced by the definition of $bal ::= Conf_{call} \ bal \ Conf_{ret}$. 
\end{proof}

To prove Theorem \ref{thm:correctcompstatefulrpc}, we use a lemma stating that the evaluation of terms in {\stateencrpc} is preserved under the let binding, as follows.
\begin{lemma}[Composition] In the stateful RPC calculus {\statefulrpc},
\label{lem:evalpreserveletstatefulrpc}
\begin{itemize}
\item If \runStatefulCSStar{M \ | \ \Delta}{V \ | \ \Delta} and it is call-return balanced\\
then \runStatefulCSStar{\llet{x}{M}{M_0} \ | \ \Delta}{\llet{x}{V}{M_0} \ | \ \Delta}, which is call-return balanced.
\item If \runStatefulCSStar{\Pi \ | \ \Delta; M}{\Pi \ | \ \Delta; V} and it is call-return balanced\\
then \runStatefulCSStar{\Pi \ | \ \Delta; \llet{x}{M}{M_0}}{\Pi \ | \ \Delta; \llet{x}{V}{M_0}}, which is call-return balanced.
\end{itemize}
\end{lemma}
\begin{proof}
This is proved by induction on the length of the evaluation in the conditions. In the base case, the length is zero, and so M is identical to V, which proves the lemma.

	In inductive cases where the evaluation takes one or more steps, the proof is done by case analysis on kinds of the first rule used in the evaluation of \runStatefulCSStar{M \ | \ \Delta}{V \ | \ \Delta} and \runStatefulCSStar{\Pi \ | \ \Delta; M}{\Pi \ | \ \Delta; V}. Hence, there are 10 cases for consideration. 
	
	i) (AppC), (AppS), (ValC), (ValS): The lemma is proved simply by induction. The resulting evaluation sequence preserves the call-return balance because it is just
the inductively obtained balanced evaluation subsequences appended in the front or in the end with $Conf_{else}$s using some of (AppC), (LetC), (ValC), (AppS), (LetS), and (ValS).
	
	ii) (LetC), (LetS): In these cases, $M$ is $\llet{x_1}{(\llet{x_2}{M_1}{M_2})}{M_3}$. These cases are proved by induction together with  Lemma \ref{lem:letbindingfirstonestatefulrpc} and Lemma \ref{lem:letbindingfirsttwostatefulrpc}. The two additional lemmas are used to uncover values which $M_1$ and $M_2$ evaluate to and then to make the evaluation of $\llet{x}{M}{M_0}$ progress until it reaches a configuration $\llet{x}{M_3\{V_{1,2}\}}{M_0}$, to which induction can be applied. The condition call-return balance is verified by $bal ::= bal bal$ because all the evaluation subsequences obtained from applying induction and the two lemmas are all call-return balanced and onely some extra configurations, which are not $Conf_{call}$ nor $Conf_{ret}$, are introduced. 
	
	iii) (Ret): The call-return balance is not satisified in the condition of this case. So, it is vacuously true. 

	iv) (Reply):  By the definition of (Reply), $M$ is a value. When $M=V$, it is trivially true with the zero evaluation step. When $M\not= V$, the evaluation sequence in the condition of this case is impossible. So, it is vacuously true.
		
	v) (Call): Let $M$ be $\llet{y}{\call{V_f}{\overline{W}}}{M_y}$ and let $\Pi$ be $\LetK{z}{M_z}$. 
	
\begin{center}
\begin{tabular}{l l l}
                 & $\Pi \ | \ \Delta; \ \llet{y}{\call{V_f}{\overline{W}}}{M_y}$ & by (Call) \\
$\RightarrowEnc$ & $\llet{z}{V_f(\overline{W})}{M_z} \ | \ \LetK{y}{M_y}\cdot\Delta$ & by the balance\\
$\RightarrowEncStar$ & $\llet{z}{\ret{V_{call}}}{M_z} \ | \ \LetK{y}{M_y}\cdot\Delta$ & by (Ret)\\
$\RightarrowEnc$ & $\Pi \ | \ \Delta; \ \llet{y}{V_{call}}{M_y}$ & by (Val)  \\
$\RightarrowEnc$ & $\Pi \ | \ \Delta; \ M_y\{v_{call}/y\}$ & by the condition \\ 
$\RightarrowEncStar$ & $\Pi \ | \ \Delta; \ V$ & 
\end{tabular}
\end{center}

It is possible to mock the evaluation steps except the last steps above with $\Pi \ | \ \Delta; \llet{x}{M}{M_0}$ because all the steps betwen $Conf_{call}$ and $Conf_{ret}$ above use only the evaluation rules that do not depend on the server stack. For the last steps of evaluation above, we prove it by induction. The call-return balance is verified by $bal ::= bal \ bal$. 

	vi) (Req): Let $M$ be $\llet{y}{\req{V_f}{\overline{W}}}{M_y}$. We do an analysis to look for a configuration as follows.
	
\begin{center}
\begin{tabular}{l l l}
                 & $\llet{y}{\req{V_f}{\overline{W}}}{M_y} \ | \  \Delta$ \\
$\RightarrowEnc$ & $\LetK{y}{M_y} \ | \ \Delta; \ \llet{r}{V_f(\overline{W})}{r}$ \\
$\RightarrowEnc$ & $Conf_1 \RightarrowEnc ... \RightarrowEnc Conf_k \RightarrowEncStar V \ | \  \Delta$
\end{tabular}
\end{center}
where either (Reply) or (Call) is used in a step, $Conf_{k-1}\RightarrowEnc Conf_k$, and neither (Reply) nor (Call) is used before the step. There should exist such a configuration.
This is because, in order to move from a server configuration $(\LetK{y}{M_y} \ | \ \Delta; \ \llet{r}{V_f(\bar{W})}{r})$ to a client configuration $(V \ | \ \Delta)$, there is no other way than using (Reply) or (Call) by the definition of the semantic rules for {\stateencrpc}. 

	For use of (Reply), it is easy to construct, after the step,
	\[
	Conf_k \RightarrowEncPlus Conf_{k+n} \RightarrowEncStar V \ | \  \Delta
	\]
	where $Conf_{k+n} \RightarrowEncStar V \ | \  \Delta$ is a condition to apply induction with. Since the subsequence from $Conf_1$ to $Conf_{k-1}$ does not contain any configurations of the form $Conf_{call}$ or $Conf_{ret}$, the condition of the lemma implies that the subsequence from $Conf_{k}$ to $V \ | \  \Delta$ is balanced. 
	
	For use of (Call), an inner induction is needed to make it progress to reach a configuration that induction of the lemma is applicable. Let $Conf_k$ be 
	\[ \LetK{y}{M_y} \ | \ \Delta; \ \llet{z}{\call{V_g}{\overline{W_{garg}}}}{M_z}
	\]
\begin{center}
\begin{tabular}{l l l}
                 & $Conf_{k}$ & by (Call) \\
$\RightarrowEnc$ & $\llet{y}{V_g(\overline{W_{garg}})}{M_y} \ | \ \LetK{z}{M_z}\cdot\Delta$ & by the balance\\
$\RightarrowEncStar$ & $\llet{y}{\ret{V_{call}}}{M_y} \ | \ \LetK{z}{M_z}\cdot\Delta$ & by (Ret)\\
$\RightarrowEnc$ & $\LetK{y}{M_y} \ | \ \Delta; \ \llet{z}{V_{call}}{M_z}$ & by (Val)  \\
$\RightarrowEnc$ & $\LetK{y}{M_y} \ | \ \Delta; \ M_z\{v_{call}/z\}$ & 
\end{tabular}
\end{center}

	The subsequence after the configuration $(\LetK{y}{M_y} \ | \ \Delta; \ \llet{z}{V_{call}}{M_z})$ all the way to $(V \ | \ \Delta)$ is balanced by the following argument. The subsequence before the configuration is constructed to be balanced. Since the whole of the evaluation sequence in the condition of the lemma is balanced, the definition of call-return balance implies that the subsequence after the configuration is balanced.
	
	We repeat the analysis again from the last configuration above until we find any use of (Reply), which allows to apply induction of the lemma. Since the steps are finite by the condition, the analysis will be terminated in finite time. 
	This finishes the proof.
\end{proof}

\begin{lemma}[Substitution]
Let $\vcomp{\lamL{a}{x}{M}}=\ccomp{\lamL{a}{x}{M}}$.
\label{lem:substitutionstatefulrpc}
\begin{itemize}
\item $\ccomp{N\{W/x\}} = \ccomp{N}\{\vcomp{W}/x\}$.
\item $\scomp{N\{W/x\}} \ K = (\scomp{N} \ K)\{\vcomp{W}/x\}$.
\end{itemize}
\end{lemma}
\begin{proof}
This lemma is true because of the definition of the compilation rules. 
\end{proof}

To prove Lemma \ref{lem:evalpreserveletstatefulrpc}, we introduce two lemma as follows.
\begin{lemma} 
\label{lem:letbindingfirstonestatefulrpc}
\begin{itemize}
\item If \runStatefulCSStar{\llet{x}{M}{M_0} \ | \ \Delta}{V \ | \ \Delta} and it is call-return balanced\\
then $\llet{x}{M}{M_0} \ | \ \Delta 
		\Rightarrow^{state*} \llet{x}{V_x}{M_0} \ | \ \Delta
		\Rightarrow^{state*} V \ | \ \Delta$ .
\item If \runStatefulCSStar{\Pi \ | \ \llet{x}{M}{M_0}}{\Pi \ | \ V} and it is call-return balanced\\
then $\Pi \ | \ \llet{x}{M}{M_0} 
		\Rightarrow^{state*} \Pi \ | \ \llet{x}{V_x}{M_0}
		\Rightarrow^{state*} \Pi \ | \ V$.
\end{itemize}
In each conclusion of the two conditional statements above, the former evaluation sequence until $M$ evaluates to $V_x$ and the latter one after that are both call-return balanced.
\end{lemma}
\begin{proof}
This is proved by induction on the length of the evaluation in the conditions.
\end{proof}

\begin{lemma}
\label{lem:letbindingfirsttwostatefulrpc}
\begin{itemize}
\item If \runStatefulCSStar{\llet{x}{M}{M_1} \ | \ \Delta}{V_1 \ | \ \Delta} and it is call-return balanced\\
then \runStatefulCSStar{\llet{x}{M}{M_2} \ | \ \Delta}{\llet{x}{V}{M_2} \ | \ \Delta} and it is also call-return balanced.
\item If \runStatefulCSStar{\Pi \ | \ \Delta; \llet{x}{M}{M_1}}{\Pi \ | \ \Delta; V_1} and it is call-return balanced\\
then \runStatefulCSStar{\Pi \ | \ \Delta; \llet{x}{M}{M_2}}{\Pi \ | \ \Delta; \llet{x}{V}{M_2}} and it is also call-return balanced.
\end{itemize}
\end{lemma}
\begin{proof}
	The lemma is proved by induction on the length of the evaluation in the conditions.
\end{proof}

The theorem of the correctness of compilation for {\statefulcs} immediately follows the theorem for {\statefulrpc}.
\begin{theorem}[Correctness of Compilation for \statefulcs] Assume a well-typed term $M$ under the locative type system. Given $\phi_\client$ and $\phi_\server$, 
\begin{itemize}
\item If $\evalRPCC{M}{V}$ then 
\runStatefulCSStar{\cconv{\ccomp{M}} \ | \ \emp}{\cconv{\ccomp{V}} \ | \ \emp}.
\end{itemize}
\end{theorem}
\begin{proof}
This theorem is proved by the same argument used in the proof of Theorem \ref{thm:correctcompstateencs}. The key idea is that the compilation rules $\cconv{-}$ preservers the structure of {\statefulrpc} terms. 
\end{proof}

\label{lastpage}
\end{document}